\documentclass[journal]{IEEEtran}

\ifCLASSINFOpdf
\else
\fi
\usepackage{animate}
\usepackage{amsmath,amssymb,amsthm}
\usepackage{upgreek}
\usepackage{mathtools}
\usepackage{bm}

\usepackage{graphicx,subfigure,epstopdf}
\usepackage{epic}
\usepackage{hyperref}
\usepackage{subfigure}
\usepackage{graphicx}
\usepackage{arydshln}
\usepackage{enumerate}
\usepackage[utf8]{inputenc}
\usepackage[english]{babel}
\usepackage{fancyhdr}
\usepackage{float}
\usepackage{lastpage}
\usepackage{color}
\usepackage{dsfont}
\usepackage{ifthen}
\usepackage{accents}
\RequirePackage{doi}
\usepackage{hyperref}
\usepackage{changes}
\usetikzlibrary{shapes,snakes}
\usepackage{verbatim}
\usepackage{ifthen}
\usepackage{pgf}
\usepackage{indentfirst}
\usepackage{todonotes}
\usepackage{makecell,booktabs}
\usepackage[load-configurations = abbreviations]{siunitx}
\usepackage{algorithm}
\usetikzlibrary{calc}
\usepackage{algorithmic}

\usetikzlibrary{shapes.geometric}

\newcount\mycount
\usetikzlibrary[topaths]
\usetikzlibrary{matrix,quotes,arrows.meta}
\definecolor{mitRed}{RGB}{117,0,20}
\definecolor{mitDargre}{RGB}{33,35,38}
\definecolor{mitSilverGray}{RGB}{139,149,158}
\definecolor{mitLSilverGray}{RGB}{184,194,204}
\definecolor{mitDPurple}{RGB}{ 62,0,107}
\definecolor{mitPurple}{RGB}{153,51,255}
\definecolor{mitLPurple}{RGB}{191,179,255}
\definecolor{mitBlue}{RGB}{25,102,255}
\definecolor{mitDGreen}{RGB}{0,77,26}
\definecolor{mitGreen}{RGB}{0,173,0}
\definecolor{mitLGreen}{RGB}{170,255,51}
\definecolor{mitPink}{RGB}{255,20,240}
\definecolor{mitLightPink}{RGB}{255,179,255}
\definecolor{mitDPink}{RGB}{117,0,98}
\definecolor{mitLightBlue}{RGB}{153,235,255}
\usepackage{tikz}
\usetikzlibrary {automata,positioning}
\usetikzlibrary{positioning}
\usetikzlibrary{arrows}
\usetikzlibrary{backgrounds,fit,shapes}
\usetikzlibrary{decorations.pathreplacing,decorations.markings}
\usetikzlibrary {arrows.meta}

\newcommand\thickbar[1]{\accentset{\rule{.4em}{.8pt}}{#1}}
\newcommand\thickbarBig[1]{\accentset{\rule{.5em}{.8pt}}{#1}}
\newcommand{\EQ}{\begin{eqnarray}}
\newcommand{\EN}{\end{eqnarray}}
\newcommand{\EQQ}{\begin{eqnarray*}}
\newcommand{\ENN}{\end{eqnarray*}}
\newcommand{\col}{\mbox{col}}

\renewcommand\thesection{\Roman{section}}
\newtheorem{theorem}{Theorem}

\newtheorem{lemma}{Lemma}
\newtheorem{remark}{Remark}
\newtheorem{definition}{Definition}

\newtheorem{problem}{Problem}

\newtheorem{assumption}{Assumption}

\newenvironment{Proof}{\noindent{\em Proof:\/}}{\hfill $\Box$\par}

\newcommand\norm[1]{\lVert#1\rVert}

\newcommand{\myr}{\color{red}}
\allowdisplaybreaks
\makeatletter
\newcommand*\bigcdot{\mathpalette\bigcdot@{.8}}
\newcommand*\bigcdot@[2]{\mathbin{\vcenter{\hbox{\scalebox{#2}{$\m@th#1\bullet$}}}}}
\makeatother 

\allowdisplaybreaks
\begin{document}
\title{Data-Driven Nonlinear Regulation: Gaussian Process Learning}
%


\author{Telema~Harry,  Martin~Guay,  Shimin~Wang, Richard D.~Braatz
\thanks{This research was supported by Mitacs and the U.S. Food and Drug Administration under the FDA BAA-22-00123 program, Award Number 75F40122C00200. \\
Telema Harry and Martin Guay are with the  Queen's University, Kingston,  ON K7L 3N6, Canada (telema.harry@queensu.ca, martin.guay@chee.queensu.ca)\\
Shimin Wang and Richard D. Braatz are with Massachusetts Institute of Technology, Cambridge, MA 02142, USA (bellewsm@mit.edu, braatz@mit.edu)
}
}

\maketitle

\begin{abstract}

This article addresses the output regulation problem for a class of nonlinear systems using a data-driven approach.
An output feedback controller is proposed that integrates a traditional control component with a data-driven learning algorithm based on Gaussian Process ($\mathcal{GP}$) regression to learn the nonlinear internal model. 
Specifically, a data-driven technique is employed to directly approximate the unknown internal model steady-state map from observed input-output data online. 
Our method does not rely on model-based observers utilized in previous studies, making it robust and suitable for systems with modelling errors and model uncertainties. 
Finally, we demonstrate through numerical examples and detailed stability analysis that, under suitable conditions, the closed-loop system remains bounded and converges to a compact set, with the size of this set decreasing as the accuracy of the data-driven model improves over time.

\end{abstract}


\section{INTRODUCTION} \label{Sect-Intro}

The ability of a control system to cancel the effect of an unknown disturbance or regulate the plant to track some reference signal is a desirable property in many engineering applications, such as cruise control, motor speed regulation, and aircraft landing and take off.
%
These problems fall under output regulation theory, encompassing both disturbance rejection and reference signal tracking, which is recognized as one of the two fundamental problems in control theory \cite{sutton2002reinforcement}.
The output regulation problem for linear systems subject to external disturbance was elegantly solved in the works \cite{FRANCIS-Wonham-1976457, Francis1977}. 
The key findings of those works are the sufficient and necessary conditions for solving the output regulation problem: ``a controller must incorporate a suitably reduplicated model of the exosystem dynamic structure with feedback of the regulated variable.'' 
This is known as the \emph{internal model principle}.

Extending the internal model principle to nonlinear systems is a challenging problem.
Unlike linear systems, the knowledge of the exosystem alone is neither sufficient nor necessary for solving the output regulation problem \cite{Byrnes-Isidori-2003,Marconi-Praly-Isidori2007, Bin_Marconi_2020}. 
It is because the steady-state dynamics and the steady-state input are combinations of the exosystem and the dynamics of the controlled system \cite{Huang_Lin_1995}.
Inspired by the linear setting, several significant contributions have been made to enhance nonlinear output regulation theory (see \cite{Byrnes-Isidori-2003, Marconi-Praly-Isidori2007, Isidori-Byrnes1990, Huang_Zhiyong2004} and references therein for more details).
In particular, \cite{Isidori-Byrnes1990} showed that the nonlinear output regulation problem is solvable by solving some sets of partial differential and algebraic equations commonly referred to as \emph{regulator equations} to obtain a nonvanishing feedforward steady-state control action. 
Inspired by the pioneering work in \cite{Isidori-Byrnes1990}, Huang \cite{Huang_Lin_1995} proposed a $k${th}-order internal model to \emph{reproduce} the solutions of the nonlinear regulator equations for uncertain nonlinear systems under the assumption that the solution of the regulator equation is polynomial in the exogenous input \cite{Huang_Zhiyong2004}. 

In subsequent years, further improvements have been made using the concepts of immersion and steady-state generators \cite{Byrnes-Isidori-2003, Marconi-Praly-Isidori2007, Huang_Zhiyong2004}. 
These approaches have led to the design of various internal models for different scenarios \cite{Huang_Isidori_Marconi_Sontag_2018}. 
The regulation equations in the cited works are described by equations whose analytical solutions are difficult to obtain even for simple problems \cite{Bin-Pauline-Marconi-2021}. 
More complex and practical scenarios, such as aerial manipulation under rotorwash, require specialized modelling and control strategies \cite{zhang2023grasping}, which challenge explicit solutions to the nonlinear regulation equation.
Furthermore, even when the regulation equations can be solved, asymptotic regulation remains a fragile property that can be lost in the presence of small perturbations in the exosystem or the plant \cite{Bin_Marconi_2023}. 
Under the assumption that the steady-state generator is polynomial in the exosystem, 
\cite{wang2024nonparametric} developed a nonparametric learning framework to learn the dynamics of the steady-state/input behaviour from internal signals.
Adaptive data-driven control strategies have been proposed in recent studies \cite{Bin-Pauline-Marconi-2021, Pauline_Bin_Marconi2020} to approximate the regulator equation solution, albeit at the cost of non-zero regulation errors. 
Especially, data-driven approaches turned out to be a powerful framework in network traffic control, which can incorporate the
model-based component facilitates the training of the ADP-based state approximator, and hence
improve the overall performance of the control system as shown in \cite{su2021adaptive}.
Likewise, SytaB \cite{yang2022sytab} illustrates the modelling challenges posed by hybrid aerial-terrestrial systems, where smooth mode transitions are difficult to capture analytically.
These complexities highlight the limitations of traditional analytical approaches and motivate data-driven methods that improve the solvability of the regulator equations by learning internal models directly from data.
%
%
A continuous-time \cite{Pauline_Bin_Marconi2020} and discrete-time \cite{Bin-Pauline-Marconi-2021} system identification technique based on a least-squares algorithm is implemented to estimate the system's internal model online. 
In general, least-squares system identification algorithms assume prior knowledge of the underlying model structure and constrain the identification problem to a finite-dimensional parameter space, which can be difficult to define for complex systems. 
The key result in \cite{Bin-Pauline-Marconi-2021,Pauline_Bin_Marconi2020} is that the asymptotic regulation performance of the data-driven regulator depends on the predictive ability of the chosen system identification algorithm.

In \cite{Gentilini-Bin-Marconi-2023}, a Gaussian Process ($\mathcal{GP}$) learning algorithm, which does not impose the finite-dimensional model constraints, is implemented with an extended observer motivated by  \cite{Bin-Pauline-Marconi-2021}. The observer design requires the computation of the time derivative of the $\mathcal{GP}$ model, thereby introducing additional numerical complexities, which can become particularly challenging for high-dimensional systems. 
The discrete-time framework in \cite{Bin-Pauline-Marconi-2021} is particularly attractive as it reduces the computational burden of continuously running the system identification algorithm. 
Building on this framework, we propose an output feedback data-driven regulator based on Gaussian process regression to learn the system's internal model online. In this control framework, the $\mathcal{GP}$ model is updated during discrete-time events and subsequently used for prediction in the continuous-time phase.
%
As a result, we formulate our proposed regulator within a hybrid dynamical system framework to effectively model the interaction between continuous-time dynamics and discrete-time events.

The proposed approach does not require complete knowledge of either the plant or the exosystem but instead relies solely on output measurements of the regulated variable. 
This enhances robustness, making the method well-suited for systems with modelling errors and parameter uncertainties.
 Furthermore, unlike the extended observer approach proposed in \cite{Bin-Pauline-Marconi-2021, Gentilini-Bin-Marconi-2023}, which requires the computation of the time derivative of the chosen system identification model, our approach eliminates this requirement, thereby simplifying numerical computations and reducing model complexity. Instead, we estimate the unknown internal model’s steady-state map directly online using a data-driven, non-parametric Gaussian Process-based internal model learning framework. Finally, this method removes the assumption that the regression model set is known a priori by developing a non-parametric representation that is not constrained to a predefined function class. 
By employing Gaussian Process (GP) regression, the proposed framework avoids the need for the polynomial steady-state generator assumption made in \cite{wang2024nonparametric}, enabling more flexible internal model learning.



This article is organized as follows. Problem formulation is presented in Section \ref{Sect: problem formulation}, some background information on internal model, and Gaussian process regression  are presented in Section \ref{Sec: preliminaries}. The Gaussian process-based regulator and the main results of this article are presented in Section \ref{Sect: Gaussian Process-Based Nonlinear Regulator }. Finally, some numerical examples to illustrate the effectiveness of a proposed algorithm are presented in Section \ref{Simulation example} and conclusions in Section \ref{Conclusion}.

\textbf{Notations:}
$\|\cdot\|$ denotes the Euclidean norm of a vector. 
If $\mathcal{A} \subset \mathds{R}^n$, then $|x|_\mathcal{A} \coloneqq \mbox{inf}_{a \mathcal{A}} | x - a| $ denotes the distance of $x \in \mathds{R}^n$ to $\mathcal{A}$. A function $\alpha: \mathds{R}_{\geq0} \rightarrow \mathds{R}_{\geq0} $ is said to be of class $\mathcal{K}$ $(\alpha \in \mathcal{K})$, if it is continuous, strictly increasing and $\alpha(0) = 0$.  A function $\gamma: \mathds{R}_{\geq0} \rightarrow \mathds{R}_{\geq0} $ is said to be of class $\mathcal{K}_\infty$, if $\gamma \in \mathcal{K}$ and, in addition   $\lim_{s \rightarrow \infty} \gamma(s) = \infty$. A function $\beta: \mathds{R}_{\geq0} \times \mathds{R}_{\geq0} \rightarrow \mathds{R}_{\geq0} $ is said to be of class $\mathcal{KL}$, if $\beta(\cdot, t) \in \mathcal{K}$ for each $t \in \mathds{R}_{\geq0} $ and, for each $s \in \mathds{R}_{\geq0}$, $\beta(s, \cdot)$ is continuous and  decreasing to zero as $t \rightarrow \infty$


\section{Problem formulation} \label{Sect: problem formulation}

This article considers the approximate output regulation problem for a class of nonlinear systems of the form
\begin{equation} \label{eqn: general nonlinear system dynamics}
    \begin{aligned}
        \dot{z} &= f_z(z,y,w,\sigma) \\
        \dot{y} & =  q(z,y, w,\sigma) + b(w,\sigma) u  
    \end{aligned}
\end{equation}
where $(z, y) \in \mathds{R}^{n_z} \times \mathds{R}$ are the system states, $u \in \mathds{R}$ is the control input, $y$ is the measured output to be regulated, and $\sigma \in \mathds{R}^{n_\sigma}$ is an uncertain parameter. 
 $w  \in \mathds{R}^{n_w}$ denotes the exogenous signal representing both the reference signal to be tracked and the disturbance signal to be rejected, and it is generated by the exosystem:
\begin{equation} \label{eqn: exosystem system}
   \begin{aligned}
        \dot{w} & = s(w, \sigma), \\
        y_0 & = h(w,\sigma),
    \end{aligned} 
\end{equation}
where $y_0 \in \mathds{R}$ is the exogenous signal's output. Define the regulated error $e \in \mathds{R}$ as 
\begin{align}\label{EQN: error system}
    e = y -  h(w,\sigma).
\end{align}
Assume that the functions $f(z,y,w,\sigma) $, $q(z,y,w,\sigma)$ are globally defined and smooth and satisfy $f(0,0,0, \sigma) = 0$, $q(0,0,0,\sigma) = 0$. The uncertain parameter $\sigma$ and exogenous signal $w$ are assumed to belong to a known compact invariant sets $\Sigma$ and $\mathds{W}$\footnote{A set $\mathds{W}$ is said to be {\em a compact invariant set} for the exosystem  $\dot{w} = s(w, \sigma)$ if it is compact and, for every initial condition $w(0) \in \mathds{W}$, the corresponding solution $w(t) \in \mathds{W}$ for all $t \geq 0$. } respectively. 


%
%
%
%
%
%
We formally define the robust practical output regulation problem composed of systems \eqref{eqn: general nonlinear system dynamics}, \eqref{eqn: exosystem system} and \eqref{EQN: error system} as follows:
\begin{problem} \label{Prob: Practical Output regulation}
Consider systems \eqref{eqn: general nonlinear system dynamics}, \eqref{eqn: exosystem system}, and \eqref{EQN: error system} and given any compact set $\mathds{W}$ and $\Sigma$ containing the origin, design an output feedback control law such that, for every initial condition $w(0) \in \mathds{W}$, and $\sigma(0) \in \Sigma$ and any states $(z(0), y(0)) \in \mathds{R}^{n_z} \times \mathds{R}$, the solution of the closed-loop system exists and is bounded for all $t \geq 0$, and $$ \lim\limits_{t\rightarrow\infty} |e (t)| \leq \varepsilon, $$
where $\varepsilon$ is a sufficiently small number. 
\end{problem}

We make some standard assumptions required for the solution of the output regulation problem.
%
%
\begin{assumption}\label{ASS: positive and lower bounded b(w,sigma)}
    The function $b(w,\sigma) $ is continuous, $b(w,\sigma)  > 0$, and lower bounded by some positive contant $b^\star$ for all $w \in \mathds{R}^w$ and all $\sigma \in \mathds{R}^{n_\sigma}$
\end{assumption}
\begin{assumption}  \label{ASS: exosystem input satisfies ISS}(Solvability of regulator equation).
    There exists a smooth functions  $\mathbf{z}^\star(w,\sigma)$ with $\mathbf{z}^\star(0,\sigma) = 0$ such that, for all $(w, \sigma) \in \mathds{W}  \times \Sigma $,
    \begin{equation}
         \frac{\partial \mathbf{z}^\star (w,\sigma)}{\partial w} s(w,\sigma) = f(\mathbf{z}^\star(w,\sigma), \mathbf{y}^\star (w,\sigma), w,\sigma)
    \end{equation}
\end{assumption}
Under Assumptions \ref{ASS: positive and lower bounded b(w,sigma)} and \ref{ASS: exosystem input satisfies ISS}, 
the ideal control input required to regulate the error $e$ to zero is given by:
\begin{align}\label{eqn: ideal steady-state control input}
    \begin{split}
        \mathbf{u}^\star( w,\sigma)   = &\ b(w,\sigma)^{-1} 
        \Big( \frac{\partial h (w,m)}{\partial w} s(w,\sigma)  \\& - q(\mathbf{z}^\star (w,\sigma),\mathbf{y}^\star (w,\sigma), w, \sigma) \Big)
    \end{split}
\end{align}
The functions $\mathbf{y}^*(w,\sigma)$, $\mathbf{z}^\star (w,\sigma)$, and $\mathbf{u}^\star(w,\sigma)$ constitute a solution to the so-called regulator equations \cite{Isidori-Byrnes1990, HuangJ-2004}.
The function $\mathbf{u}^\star(w,\sigma)$ provides the necessary feedforward control action to maintain zero regulation error when the system is at steady state. Any regulator capable of generating the feedforward control action $\mathbf{u}^\star(w,\sigma)$ is said to possess the \emph{``internal model property"} \cite{Isidori-Byrnes1990}. 
Although the expression of $\mathbf{u}^\star(w,\sigma)$ can be easily derived, it cannot be used for feedback control design, as it contains the exogenous signal $w$ and the unknown parameter $\sigma$.
In practice, an internal model is designed to reproduce $\mathbf{u}^\star(w,\sigma)$ asymptotically.

\section{Preliminaries} \label{Sec: preliminaries}

\subsection{Hybrid Dynamical Systems}

This article proposes a regulator described by a hybrid system using the formalism of \cite{goedel2012hybrid}. We consider a hybrid system $\mathcal{H}$ with state $x \in \mathds{X}$ and input $u \in \mathcal{U}$ of the form
%
%
\begin{equation} \label{EQN: classic hybrid system}
\mathcal{H}: \quad
    \begin{cases} 
       \quad (x,u) \in \mathcal{C}, \: \:\:\:\:\:\:\: \:\:\:  \dot{x} = f(x,u)  \\
      \quad (x,u) \in \mathcal{D}, \: \:\:\:\:\:\:\: \: \:    x^+ = g(x,u)   
   \end{cases}
 \end{equation}
where $\mathcal{C}$ denotes the \emph{flow set} and  $\mathcal{D}$ denotes the \emph{jump set}. The flow equation describes the continuous evolution of the state $x$. The jump equation describes the discontinuous evolution of the state. The vector fields $f$ and $g$ are assumed to be continuous on $\mathcal{C}$ and $\mathcal{D}$, respectively.  The solutions of system \eqref{EQN: classic hybrid system} are defined on a hybrid time domain. We recall some definitions related to the hybrid formulation from \cite{goedel2012hybrid}.
\begin{definition} (Hybrid time domains)
A subset $E \subset \mathds{R}_{\geq 0} \times \mathds{N}$ is a {\em hybrid time domain} if, for all $(T,J) \in E$, $$E \cap ([0,T] \times \{0,1,\cdots{}, J\}) = \bigcup_{j = 0}^{J-1}([t_j, t_{j+1}], j) $$ for some finite sequence of times $0 = t_0 \leq t_1 \leq \cdots \leq t_J$.   
\end{definition}
\begin{definition} (Hybrid arc)
A function $\phi: E \rightarrow \mathds{R}^n$ is a {\em hybrid arc} if $E$ is a hybrid time domain and if, for each $j \in \mathds{N}$, the function $t \mapsto \phi(t,j)$ is locally absolutely continuous on the internal $I^j = \{t: \; (t,j) \in E \}$   
\end{definition}
%
%
\begin{definition}\label{Definition: Local pre-asymptotic stability}(Local pre-asymptotic stability (LpAS)) Let $\mathcal{H}$ be a hybrid system in $\mathds{R}^n$. A compact set $\mathcal{A} \subset \mathds{R}^n$ is said to be
\begin{itemize}
    \item {\em stable for $\mathcal{H}$} if, for every $\varepsilon > 0$, there exists $\delta > 0$ such that every solution $\phi$ to $\mathcal{H}$  with $|\phi(0,0)|_\mathcal{A} \leq \delta$ satisfies $|\phi(t,j)|_\mathcal{A} \leq \varepsilon$ for all $(t,j) \in \text{dom} \; \phi$;
    \item {\em locally pre-attractive for $\mathcal{H}$} if there exists $\mu > 0$ such that every solution $\phi$ to $\mathcal{H}$ with $|\phi(0,0)|_\mathcal{A} \leq \mu$ is bounded and, if $\phi$ is complete, then also $\lim_{t+j \rightarrow \infty} |\phi(t,j)|_\mathcal{A} = 0$;
    \item {\em locally pre-asymptotically stable for $\mathcal{H}$} if is both stable an locally pre-attractive for $\mathcal{H}$.
\end{itemize}  
\end{definition}

%

%

\subsection{Internal Model and Augmented System}\label{Subsection: Linear Internal Model and Augmented System}

Motivated by \cite{Marconi-Praly-Isidori2007, Huang_Zhiyong2004},  construct a generic internal model of the form 
\begin{equation}\label{EQN: General Internal Model}
    \dot{\eta} = M \eta + N u, \quad \eta \in \mathds{R}^{n_\eta}
\end{equation}
for the nonlinear system composed of \eqref{eqn: general nonlinear system dynamics}, \eqref{eqn: exosystem system}, and \eqref{EQN: error system}
where $(M,N) \in \mathds{R}^{n_\eta \times n_\eta } \times \mathds{R}^{n_\eta  \times 1 } $ is a controllable pair, with $M$ being a Hurwitz matrix, and $n_\eta  = 2(n_w + n_z + 1)$ according to \cite{Marconi-Praly-Isidori2007}. As shown in \cite{Marconi-Praly-Isidori2007, Marconi_Praly-2008}, the internal model \eqref{EQN: General Internal Model} can be derived using the results from \cite{Kreisselmeier_Engel-2003}.
%
Motivated by \cite{Shimin_Martin_Chen_Richard_2024}, let
\begin{equation} \label{EQN: observation mapping}
\begin{aligned}
    \bm{\eta}^\star(w(t), \sigma) & = \int_{-\infty}^{t} \exp({M(t-\tau})) N \mathbf{u}^\star(w(\tau), \sigma) \mathrm{d} \tau \\
    \mathbf{u}^\star(w(t),\sigma) & = \upvarpi(\bm{\eta}^\star(w(t), \sigma)), \quad \bm{\eta} \in \mathds{R}^{n_\eta }
\end{aligned}
\end{equation}
be defined along the system trajectory associated with the initial state; then  \eqref{EQN: observation mapping} satisfies the differential equations
%
%
%
\begin{align}
   \frac{\mathrm{d}(\bm{\eta}^\star(w(t),\sigma)) }{\mathrm{d} t}& = M \bm{\eta}^\star(w(t),\sigma) + N \mathbf{u}^\star(w(t),\sigma) \\
    \mathbf{u}^\star(w(t),\sigma) & = \upvarpi(\bm{\eta}^\star(w(t),\sigma)) \label{EQN: Internal model nonlinear map}
\end{align}
where $\upvarpi(\bm{\eta}^\star(w(t),\sigma))$ is an unknown continuous nonlinear mapping \cite{Marconi-Praly-Isidori2007}. 
For convenience, let $\bm{\eta}^\star(t) \equiv \bm{\eta}^\star(w(t),\sigma) $. 
%
%
Define the ``error variables''
\begin{equation} \label{Eqn: coordinate transformation-data driven output regulation}
    \begin{aligned}
        \thickbar{z} & = z - \mathbf{z}^\star(w,\sigma), \\
        \thickbar{\eta} & = \eta - \bm{\eta}^\star  - b^{-1}(w,\sigma) Ne, 
    \end{aligned}
\end{equation}
which describe the deviations of the state variables from their ideal steady state.
%
%
The time derivative of the error variables can be expressed as
\begin{subequations} \label{EQN: augmented system}
    \begin{align}
        \dot{\thickbar{z}} &= \thickbar{f}_z(\thickbar{z}, e, w, \sigma),\\
        \dot{\thickbar{\eta}} & = M \thickbar{\eta} + \thickbar{r}(\thickbar{z},e,w, \sigma), \\
        \dot{e} & = \thickbar{q}(\thickbar{z}, \thickbar{\eta},e,w, \sigma) + b(w, \sigma) (u -  \mathbf{u}^\star(w, \sigma)   ),
    \end{align}
\end{subequations}
where
\begin{align*} 
    \begin{split}
\thickbar{f}_z(\thickbar{z},e,w, \sigma) = &\ f_z(\thickbar{z} + \mathbf{z}^\star, e + h(w,\sigma), w, \sigma) \\
        & - f_z(\mathbf{z}^\star,h(w,\sigma), w, \sigma),\\
       \thickbar{r}(\Bar{z},e,w, \sigma)  = &\ b(w, \sigma)^{-1}(M N e - N\thickbar{q}(\thickbar{z},e, w, \sigma) ) \\
       & - \frac{\mbox{d} b(w, \sigma)^{-1} }{\mbox{d}t}Ne, 
    \end{split} \\
   \thickbar{q}(\thickbar{z}, \thickbar{\eta}, e,w, \sigma) = &\  q(\thickbar{z} + \mathbf{z}^\star ,e +h(w,\sigma), w, \sigma)   \\
   & -   q(\mathbf{z}^\star,h(w,\sigma), w, \sigma).
\end{align*}
System \eqref{EQN: augmented system} is called the augmented system. It can be shown that, for any $w, \sigma \in \mathds{W}  \times \Sigma$, 
$$ \thickbar{f}_z(0,0,0, \sigma) = 0, \quad \thickbar{r}(0,0,0, \sigma) = 0, \quad \thickbar{q}(0, 0, 0,0, \sigma) = 0. $$
Therefore, the augmented system \eqref{EQN: augmented system} possesses a stable equilibrium point at the origin, where the regulated variable $e$ is zero.

\subsection{Gaussian Process Regression Identifier }\label{Sec: Gaussian Process Regression Identifier}

Equation \eqref{EQN: Internal model nonlinear map} can be interpreted as a regression model that relates $\mathbf{u}^\star$  to $\bm{\eta}^\star$. This interpretation enables the application of regression techniques to reconstruct the unknown nonlinear mapping $\upvarpi$ from a suitable input–output data pairs $$\{\bm{\mathfrak{N}_r}, \bm{\mathfrak{U}_r}  \}_{r \in \mathds{N}},$$ with $$\bm{\mathfrak{N}_r} = \{\bm{\eta}^\star_1,\dots,\bm{\eta}^\star_r \}\;\;\;\textnormal{and}\;\;\; \bm{\mathfrak{U}_r} = \{\mathbf{u}^\star_1,\dots,\mathbf{u}^\star_r \}.$$
To develop the regression model, assume access to samples of two \emph{proxy variables}, which serve as a perturbed version of $\bm{\eta}^\star$ and $\mathbf{u}^\star$, since the exact quantities are unknown. 
As shown in \cite{Bin-Pauline-Marconi-2021, Pauline_Bin_Marconi2020}, $\eta$ and $u$ can serve as suitable proxies. These proxies enable us to approximate the unknown relationship while preserving the structure of the problem.

We propose a Gaussian Process $(\mathcal{GP})$ model to approximate the unknown function $\upvarpi$.
%
A Gaussian Process $\mathcal{GP}$ is a probabilistic, non-parametric approach for modelling complex functions based on observed data. Formally, a 
$\mathcal{GP}$ is defined as a collection of random variables, any finite subset of which follows a joint Gaussian distribution \cite{Rasmussen_Williams2006}. Unlike deterministic models that yield a single ``best fit'' to the data, a 
$\mathcal{GP}$ provides a distribution over functions, making it particularly well-suited for applications requiring uncertainty quantification.

A $\mathcal{GP}$ of a real process $\upvarpi(\bm{\eta})$ is completely characterized by its mean function $m(\bm{\eta})$ and its covariance function $K(\bm{\eta}, \bm{\eta'}) $ which captures the correlation between any two inputs $\bm{\eta}$ and $\bm{\eta'}$. The $\mathcal{GP}$  is formally expressed as $$\upvarpi(\bm{\eta}) \sim \mathcal{GP}(m(\bm{\eta}), K(\bm{\eta}, \bm{\eta'}))$$
where 
\begin{equation}
    \begin{aligned}
        m({\eta}) &= \mathds{E}[\upvarpi(\bm{\eta})] \\
       K(\bm{\eta}, \bm{\eta'}) & = \mathds{E}[(\upvarpi(\bm{\eta}) - \mu(\bm{\eta}))(\upvarpi(\bm{\eta'}) - \mu(\bm{\eta'})) ].
    \end{aligned}
\end{equation}
The $\mathcal{GP}$ approximation of the unknown nonlinear map $\upvarpi(\cdot)$ is a two-step process, comprising of \emph{learning} and \emph{prediction}. In the \textbf{learning phase}, input–output datasets generated by a continuous-time system are used to train the $\mathcal{GP}$ model. 
The training data $\mathcal{D}_S \coloneqq \{\bm{\eta}, \bm{u}\}_{i\in \mathds{N}}$ consists of the input dataset $$\bm{\eta} = \col(\eta(t_1), \eta(t_2), \cdots{}, \eta(t_N)) \in \mathds{R}^{n_\eta} $$ with the corresponding output dataset $$\bm{u} =\col(u(t_1), u(t_2), \cdots{}, u(t_N))\in \mathds{R}^{N}. $$ 
%

As typical in the literature, assume a zero mean $\mathcal{GP}$, without loss of generality. In this work, a squared exponential (SE) kernel is used:
\begin{equation} \label{EQN: squared exponential kernel function}
    K(\bm{\eta},\bm{\eta'}) = \sigma_f^2 \exp\!{\left( - \frac{ \| \bm{\eta} - \bm{\eta}' \|^2}{ 2 \ell ^2}\right)},
\end{equation}
where $\sigma_f^2 \in \mathds{R}_{\geq 0}$ is the variance parameter and $\ell_i \in \mathds{R}_{\geq 0}, \; i = 1, \cdots{}, n$ are the length scales. $\sigma_f^2$ and $\ell$ are commonly referred to as hyperparameters. The hyperparameters are obtained from a maximum likelihood argument following Bayesian principles. 

In the \textbf{prediction phase}, the trained $\mathcal{GP}$ model is used to estimate the unknown function $\upvarpi$ at new, unseen input points $$\bm{\eta'} = \col(\eta'(t_i), \cdots{},\eta'(t_J)) \in \mathds{R}^{n_\eta}. $$ The prediction is performed by computing the posterior distribution conditioned on the training data $\mathcal{D}_S$. The posterior mean and variance of the predicted outputs $\bm{u}'$ are given by
%
    \begin{align}
         \mu (\bm{\eta'}) = &\ K(\bm{\eta}, \bm{\eta'})^\text{T} [ K(\bm{\eta},\bm{\eta}) + \sigma_n^2 I]^{-1} \bm{u} \label{EQN: GP posterior mean} \\
        \begin{split}
            \sigma^2(\bm{\eta'}) = &\ K(\bm{\bm{\eta'}, \bm{\eta'}}) - K(\bm{\eta}, \bm{\eta'})^\text{T} [ K(\bm{\eta},\bm{\eta}) + \sigma_n^2 I]^{-1} \\& \times  K(\bm{\bm{\eta}, \bm{\eta'}}) 
        \end{split}\label{EQN: GP variance}
    \end{align}
where $K(\bm{\eta},\bm{\eta})$ denotes the covariance matrix of the training inputs, $K(\bm{\eta'}, \bm{\eta})$ is the covariance matrix between the test inputs and the training inputs, and $\sigma^2 I$ accounts for the Gaussian noise $\varepsilon$ in the observation. 



The $\mathcal{GP}$ model incrementally learns the unknown nonlinear mapping $\upvarpi$ from the training dataset $\mathcal{D}_S$.
Its learning rate is characterized by a learning curve that relates the Bayesian generalization error $\epsilon_D$
  to the size of the training dataset, and it is independent of the test points \cite{Francesco-thesis-1998}. For a noise-free output, the Bayesian generalization error $\epsilon_D$
  at a given test point $\upvarpi(\bm{\eta}^\star)$
  is defined as
\begin{equation}\label{EQN: Bayesian generalization error}
\epsilon_D = \left(\mu(\bm{\eta}^\star) - \upvarpi(\bm{\eta}^\star)\right)^2.
\end{equation}
For a noise-free output, it can be readily shown that the posterior variance, as given by \eqref{EQN: GP variance}, is equivalent to the expected value of the Bayesian generalization error $\epsilon_D$ for a given a training dataset $\mathcal{D}_S$ \cite{Francesco-thesis-1998}.


\begin{assumption}\label{Ass: Unknown function is a Gaussian Process}
    The unknown function $\upvarpi(\bm{\eta^\star})$ is a sample drawn from the Gaussian process $(\mathcal{GP})$ with zero mean and covariance (kernel) function $K(\bm{\eta}, \bm{(\eta')}$, i.e.,
    $$\upvarpi(\bm{\eta}) \sim \mathcal{GP}(0, K(\bm{\eta}, \bm{\eta'})).$$
\end{assumption}


\begin{assumption}\label{Ass: Bounded RKHS}
    The unknown function $\upvarpi$ has a bounded norm  under the \textbf{RKHS} generated by the kernel $K(\bm{\eta},\bm{\eta'})$.  
\end{assumption}

\begin{assumption}\label{Ass: Gaussian Process - Lipschitz}
The covariance kernel $K(\bm{\eta},\bm{\eta'})$ is smooth and Lipschitz continuous with a constant $\mathcal{L}_k$. 

\end{assumption}
Assumptions \ref{Ass: Bounded RKHS} and \ref{Ass: Gaussian Process - Lipschitz} are standard in the literature, see \cite{ Lederer-Umlauft-Sandra-2019, Gentilini-Bin-Marconi-2022}. Assumption \ref{Ass: Bounded RKHS} ensures that the unknown function is not discontinuous.  Moreover, Assumption \ref{Ass: Gaussian Process - Lipschitz} is inherently satisfied by most commonly used covariance functions, such as the squared exponential kernel defined in \eqref{EQN: squared exponential kernel function}. 

Next, we present key results from the literature that will serve as the basis for establishing our results.

\begin{lemma} \label{LEMMA: Upper bounds for GP}\cite[Thm.\ 9]{Lederer-Umlauft-Sandra-2019}
    Consider a zero mean Gaussian process defined through the continuous covariance kernel $K(\cdot,\cdot)$ with Lipschitz constant $L_k$ on the compact set $\mathds{X}$. Furthermore, consider a continuous unknown function $f : \mathds{X} \to \mathbb{R}$ with Lipschitz constant $L_f$ and $r \in \mathbb{N}$ observations $y^{(i)}$ satisfying Assumptions \ref{Ass: Unknown function is a Gaussian Process} and \ref{Ass: Bounded RKHS}. Then, the posterior mean $\mu(\cdot)$ and posterior standard deviation $\sigma^2(\cdot)$ of a Gaussian process conditioned on the training data $\{(x^{(i)}, y^{(i)})\}_{i=1}^r$ are continuous with Lipschitz constants $L_{\mu_r}$ and $L_{\sigma^2_r}$ on $\mathds{X}$, respectively, where
\begin{align*}
    L_{\mu_r} &\leq L_k \sqrt{r} \left\| \left( \bm{K}(X_r, X_r) + \sigma_n^2 I_r \right)^{-1} \bm{y}_r \right\|, \\
    L_{\sigma^2_r} &\leq 2\rho L_k \!\left( 1 + r \left\| \left( \bm{K}(X_r, X_r) + \sigma_n^2 I_r \right)^{-1} \right\| \max_{x, x' \in \mathcal{X}} k(x, x') \right)
\end{align*}
Moreover, pick $\delta \in (0,1)$, $\rho \in \mathbb{R}_+$ and set
\begin{align*}
    \beta(\rho) &= 2 \log\!\left( \frac{M(\rho, \mathds{X})}{\delta} \right), \\
    \gamma(\rho) &= (L_{\mu_r} + L_f)\rho + \sqrt{\beta(\rho)} L_{\sigma^2_r} \rho,
\end{align*}
where $M(\rho, \mathcal{X})$ denotes the $\rho$-covering number\footnote{The minimum number satisfying $\min_{x\in \mathds{X}} \max_{x' \in \mathcal{D_S}} \|x - x'\| \leq \rho $} of $\mathds{X}$, i.e., the minimum number such that there exists a set $\mathds{X}_\rho$ satisfying $|\mathds{X}_\rho| = M(\rho, \mathds{X})$ and $\forall x \in \mathds{X}$ there exists $x' \in \mathds{X}_\rho$ with $\|x - x'\| \leq \rho$. Then the bound
\begin{equation*}
    \left| f(x) - \mu_r(x) \right| \leq \sqrt{\beta(\rho)} \sigma_r(x) + \gamma(\rho), \ \forall x \in \mathds{X} 
\end{equation*}
holds with probability of at least $1 - \delta$
\end{lemma}

\section{Gaussian Process-Based Nonlinear Regulator}\label{Sect: Gaussian Process-Based Nonlinear Regulator }

\begin{figure*}
    \centering
    \begin{tikzpicture}[
    box/.style={rectangle, draw, rounded corners, minimum height=2em, minimum width=5em, align=center, font=\small, fill=blue!6},
    arrow/.style={-Stealth, thick},
    node distance=1.6cm and 2.2cm
]

\draw[ thick, loosely dotted,fill=mitRed!5,label={below:Discrete-Time Phase}] (-8.0, 1.25)--(2.2,1.25) -- (2.2,-5.1) -- (-8.0,-5.1) -- (-8.0, 1.25) ;

\draw[ thick, loosely  dotted,fill=mitGreen!5] (-2.3, 2.75)--(6.7,2.75) -- (6.7,-3.6) -- (-2.3,-3.6) -- (-2.3, 2.75) ;

\node[box] (system) {{\color{mitDPink}\bf System Dynamics}\\
{\bf $\begin{aligned}
                 \dot{z} &= f(z,y,w,\sigma) \\
                \dot{y} & =  q(z,y, \varphi) + b(z,y, \varphi) u  \\
                \dot{\eta} & = M \eta + N u \\
                e & = y - h(w,m)
            \end{aligned}$}};
\node[box, right = 1.3cm of system.-20] (eta) {{\color{mitDPink}\bf Internal State}\\ {\bf\(\eta(t),~e(t)\)}};
\node[box, right=1.3cm of system.25] (exosystem) {{\color{mitDPink}\bf Exosystem}\\ {\bf\(\dot{w}  = s(w,\sigma) \)}};

\node[box, below  =1.05cm of eta] (u) {\bf {\color{mitDPink}\bf Control Action}\\ \(u(t) = \hat{\upvarpi}_k(\eta(t)) + \rho(e(t))\)};
\node[box, below=0.8cm of system] (measure) {\bf{\color{mitDPink}\bf Measurement}\\ \(\eta(t_k),~u(t_k)\)};
\node[box, left=1.5cm of measure] (dataset) {\bf{\color{mitDPink}\bf Sliding Window}\\ \(D_k = \{(\eta(t_i), u(t_i))\}_{i=\max(0, k- \mathcal{P}+1)}^k\)};
\node[box, above=1.6cm of dataset] (gp) {{\color{mitDPink}\bf GP Model}\\ \(\hat{\upvarpi}_k(\cdot)\)};

\node[draw, circle, fill, minimum size=2pt, inner sep=0pt] (dot1) at ($(system.-20) + (0.5cm, 0)$) {};
\draw[arrow] (system.-20) -- (eta) ;
\draw[arrow] (eta) -- (u) node[midway, right, xshift=2mm] {\small {\color{mitDPink}\bf Input}} ;
\draw[arrow,dashed] (dot1) |- (measure.15);
\draw[arrow,dashed] (u) -- (measure.-15) node[midway, above] {\small {\color{mitDPink}\bf at \(t_k\)}};

\draw[arrow] (dataset) -- (gp) node[midway, left] {\small {\color{mitDPink}\bf Learning}};
\draw[arrow] (measure) -- (dataset) node[midway, above] {\small {\color{mitDPink}\bf Update}};
\draw[arrow] (exosystem)--(system.25);
\draw[arrow] (u.east) --++ ( 0.2,0) --++ (0,4.5)  -|  (system.north);


\draw[arrow] (gp.west) --++ ( -1.8,0) --++ (0,-4.35)  -|  (u.south);



\node[loosely dotted,above right =1cm and -0.4cm of system, font=\bfseries] (cont) {\color{mitDPink}\bf Continuous Phase: \(t \in [t_k, t_{k+1})\)};
\node[below left =1.7cm and -4.05cm of dataset, font=\bfseries] (disc) {\color{mitDPink}\bf Discrete Update: at \(t = t_k\)};

\node[above right =1.7cm and -2.7cm of system,] (cont) {\Large\color{mitDGreen}\bf Continuous-Time Phase};
\node[below right =2.3cm and -4cm of dataset] (disc) {\Large\color{mitDPurple}\bf Discrete-Time Phase};


\end{tikzpicture}
\caption{Hybrid Framework for a Closed-Loop Control System with a Gaussian Process-Based Nonlinear Regulator: The system operates in a continuous phase (\(t \in [t_k, t_{k+1})\)) where the GP model predicts the control action \(u(t) = \hat{\upvarpi}_k(\eta(t)) + \rho(e)\), and a discrete phase (at \(t = t_k\)) where the GP model is updated using a sliding window of the last \(P\) measurements of \(\eta(t)\) and \(u(t)\).}
    \label{Fig: GP-based Regulator hybrid Block diagram}
\end{figure*}
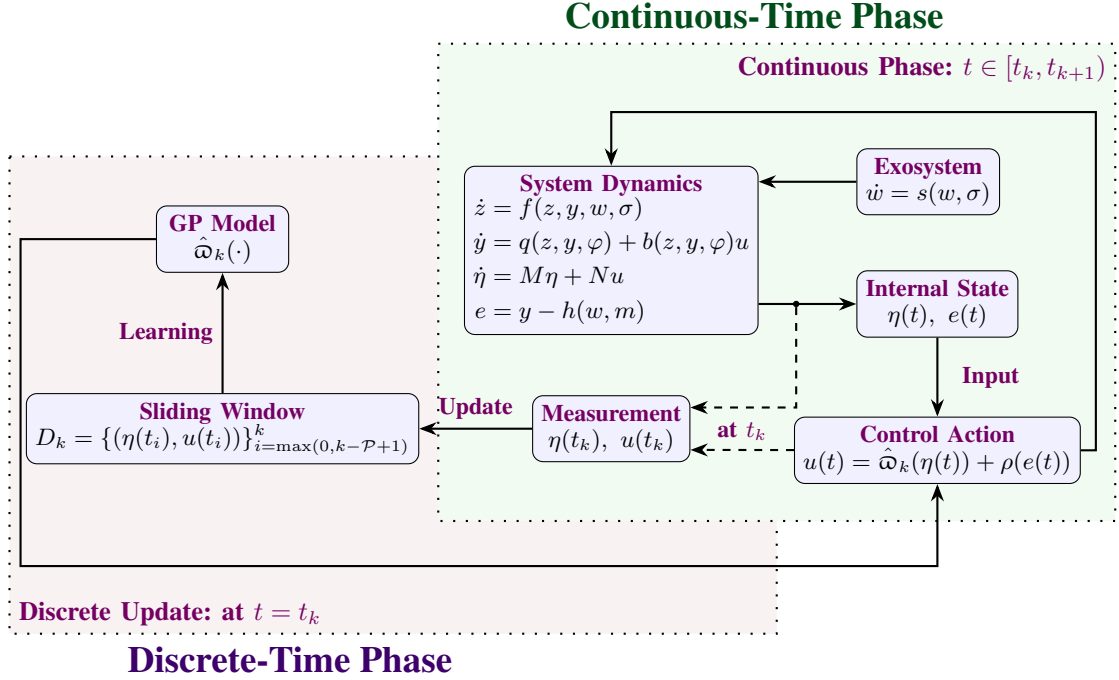

%
%
We propose a data-driven regulator that iteratively utilizes online measurements of the internal model state $(\eta)$ and the corresponding control action $(u)$ to approximate the unknown continuous function $\upvarpi(\cdot)$ using a Gaussian Process $\mathcal{(GP)}$  model. A sliding window of size $P$ is employed to maintain a manageable and up-to-date dataset $\mathcal{D_S}$. The $\mathcal{(GP)}$  model is updated at discrete-time instances and subsequently used for prediction during continuous-time evolution, as illustrated in Figure~\ref{Fig: GP-based Regulator hybrid Block diagram}. This $\mathcal{GP}$-based internal model learning regulator is formulated within a hybrid system framework as
\begin{subequations} \label{EQN:Gaussian Process-Based Nonlinear Regulator} \begin{align} & 
\begin{cases} \dot{\tau} = 1, \\ 
\dot{\eta} = M\eta + N \hat{u}, \\ 
\dot{\hat{u}} = 0, 
\end{cases} \nonumber \\
&\quad C \coloneqq [0, T] \times \mathbb{R}^{n_\eta} \times \Theta  \label{eqn:observer flow set1} \\ 
& \begin{cases} \tau^+ = 0, \\ 
\eta^+ = \eta, \\ 
\hat{u}^+ = \hat{\upvarpi}(\mathcal{D_S}), 
\end{cases} \nonumber \\
& \quad D \coloneqq \{T\} \times \mathbb{R}^{n_\eta} \times \Theta  \label{eqn:observer jump set1} 
\end{align} 
\end{subequations}
where the training dataset obtained by the sliding window of size $\mathcal{P}$ is given by
$$\mathcal{D_S}^+ = ( \mathcal{D_S} \cup \{(\bm{\eta}, \hat{u})\}) \setminus \{(\eta(t_{k-\mathcal{P}}), u(t_{k-\mathcal{P}}))\}.$$ The $\mathcal{GP}$ model is updated at each discrete time event based on a sliding window of size $\mathcal{P}$, 
%
where $\hat{\upvarpi}(\mathcal{D_S}) \equiv \mathcal{GP}(\mu(\bm{\eta}), K(\bm{\eta}, \bm{\eta'}))$.
The notation $\hat{\upvarpi(\mathcal{D_S})}$ is a mild abuse of notation, used to emphasize the dependence of the learned model on the specific training data $\mathcal{D_S}$.

The feedback controller is given by
%
    \begin{align}
   u & =-k_p \rho(e) e +\mbox{sat}( \mu(\eta)) \label{eqn: control-law regulator equation}  \\
   \mu(\eta) & =  \hat{\upvarpi}(\eta)
\end{align}
where $k_p \in \mathds{R}_{\geq 0}$ and $\rho(e)$ are the controller gain and smooth function to be designed, respectively.\footnote{Recall that the $\mathcal{GP}$ model, denoted by $\hat{\upvarpi}(\mathcal{D_S})$, is characterized by its predictive mean function $\mu(\eta)$, where $\eta \in \mathbb{R}^{n_\eta}$ is the input feature vector at which the prediction is evaluated.}
%

%
%
The regulator is formulated using a hybrid system framework to manage the control process effectively. It includes a hybrid clock $\tau \in [0,T]$, which ensures that the system transitions between continuous and discrete phases at precise intervals $T$ by resetting $\tau$ to zero when it reaches a predefined threshold $T$. 

The internal model state $\eta \in \mathds{R}^{n_\eta}$ evolves continuously according to the dynamical system given in \eqref{eqn:observer flow set1} and its values are used to update the training dataset $\mathcal{D_S}$ at discrete time instances. 
Let $\chi \coloneqq \col(\tau, \thickbar{z}, \thickbar{\eta},e, \hat{u})$.
The coupling of the augmented system \eqref{EQN: augmented system}, regulator \eqref{EQN:Gaussian Process-Based Nonlinear Regulator} and control law \eqref{eqn: control-law regulator equation} is given as
%
%
\begin{subequations} \label{EQN: close loop hybrid regulator equation - transformation}
\begin{align}
    f(\chi) \coloneqq
    &
    \begin{cases}
        \dot{\tau} = 1 \\
        \dot{\bar{z}} = \bar{f}_z(\bar{z}, e, w, \sigma) \\
        \dot{\bar{\eta}} = M \bar{\eta} + \bar{r}(\bar{z}, e, w, \sigma) \\
        \dot{e} = \bar{q}(\bar{z}, \bar{\eta}, e, w, \sigma) + b(w, \sigma) \times   \\
       \hspace{0.25in}\big(\!-k_p \rho(e) e+ \mathrm{sat}(\mu(\eta)) - \upvarpi(\bm{\eta}^\star) \big) \\
        \dot{\hat{u}} = 0 
    \end{cases}
    \label{eqn: observer flow set1 - transformation} \\
    & \mathcal{C} \coloneqq [0, T] \times \mathds{R}^{n_z} \times \mathds{R}^{n_\eta} \times \mathds{R} \times \Theta
    \nonumber \\
    %
    g(\chi) \coloneqq
    &
    \begin{cases}
        \tau^+ = 0 \\
        \bar{z}^+ = \bar{z} \\ 
        \bar{\eta}^+ = \bar{\eta} \\
        e^+ = e \\
        \hat{u}^+ = \hat{\upvarpi}(\mathcal{D_S})
    \end{cases}
    \label{eqn:observer jump set1-transformation} \\
    & \mathcal{D} \coloneqq \{T\} \times \mathds{R}^{n_z} \times \mathds{R}^{n_\eta} \times \mathds{R} \times \Theta
    \nonumber
\end{align}
\end{subequations}
where the continuous dynamics of the nonlinear system is governed by the function $f(\chi)$ on  set $\mathcal{C}$ and the discrete dynamics on set $\mathcal{D}$ are governed by the function $g(\chi)$.

%
%
%
%
We present our first result for the $(\thickbar{z}, \thickbar{\eta})$--subsystem. 

\begin{lemma}\label{COR: ISS results for z dynamics system}
    Consider the $\thickbarBig{Z} \coloneqq \textnormal{\col}(\thickbar{z}, \thickbar{\eta})$ subsystem in \eqref{EQN: close loop hybrid regulator equation - transformation}, there exist a smooth function $V_1(\thickbarBig{Z}): \mathds{R}^{n_z} \times \mathds{R}^{n_\eta} \rightarrow \mathds{R}_{\geq 0} $, $\Uppsi(\thickbarBig{Z}) > 0$ and $\thickbarBig{\Uppsi}(\thickbarBig{Z}) > 0$, a smooth positive definite function $\gamma_r(e) > 0$, and some class $\mathcal{K}_\infty$ functions $\underbar{$\alpha$}_1$, $\thickbar{\alpha}_1$ such that
 \begin{subequations} \label{EQN: nz subsystem lemma}
    \begin{align}
\underbar{$\alpha$}_1(\thickbarBig{Z}) \leq   V_1(\thickbarBig{Z}) \leq&\ \thickbar{\alpha}_1(\thickbarBig{Z}) \label{EQN: Lyapunov condition for Lemma} \\
        \begin{split}
            \langle \nabla  V_1(\thickbarBig{Z}),  f_z(\thickbarBig{Z}) \rangle \leq&  -\Uppsi(\thickbarBig{Z})\thickbarBig{Z}^2  + \gamma_r(e)e^2 
        \end{split} \label{EQN: zn subsystem flow equation}
    \end{align}
 \end{subequations}  
%
\end{lemma}
The proof is in Section  \ref{Appendix: Proof of Corollary - ISS of Z subsystem}.

As in \cite{Huang_Zhiyong2004,Shimin_Guay_Richard-2024}, we focus on solving the robust stabilization problem of the augmented system \eqref{EQN: close loop hybrid regulator equation - transformation} with the controller \eqref{eqn: control-law regulator equation}.
We now present the main result of this article with the following theorem.


\begin{theorem}
Under Assumptions \ref{ASS: positive and lower bounded b(w,sigma)}--\ref{Ass: Gaussian Process - Lipschitz}, the closed-loop system obtained by the interconnection of systems \eqref{eqn: general nonlinear system dynamics}, \eqref{eqn: exosystem system} and \eqref{EQN: error system} with the Gaussian process-based regulator \eqref{EQN:Gaussian Process-Based Nonlinear Regulator} and control law \eqref{eqn: control-law regulator equation} exhibits the following stability property: there exists a sufficiently large positive constant $k_p$ and a positive smooth function $\rho(e)$ such that any solution $\phi$ of the hybrid system $\mathcal{H}_u$ defined by \eqref{EQN: close loop hybrid regulator equation - transformation} originating from the compact set $\mathcal{A} \coloneqq \{0, T\} \times \mathds{R}^{n_z} \times \mathds{R}^{n_\eta} \times \mathds{E} \times \Theta$, $(\tau, \thickbar{z}, \thickbar{\eta},e, \hat{u})$ is bounded for all $(t,j) > 0 \times 0$ and 
the solution  $\phi$ satisfies $|\phi(t,j)|_{\mathcal{A}} \leq  \varepsilon_\mu $  as (t, j) $\rightarrow \infty \times \infty$ where $\varepsilon_\mu$ is a sufficiently small number centred around zero. 

\end{theorem}

\begin{Proof}
We pose the Lyapunov function: 
\begin{equation}\label{EQN: Main Lyapunov function}
    V(\chi) =\exp(T-\tau) (V_1(\thickbarBig{Z}) + e^2  + (\hat{u} - \mathbf{u^\star})^2 )
\end{equation}
where $u^\star$ denotes the ideal feedforward control input and $\hat{u}$ is its approximation. 
%
%
We begin by demonstrating that $V(\chi)$ satisfies the inequalities
 $$\alpha_5(\norm{ \chi}) \leq  V(\norm{\chi}) \leq \alpha_6(\norm{\chi}) $$ for some class $\mathcal{K}_\infty$ functions $\alpha_5(\cdot)$ and $\alpha_6(\cdot)$. Specifically, $V(\norm{\chi})$ satisfies the inequalities
\begin{equation}\label{eq:V_bounds}
\begin{aligned}
    V(\chi) \geq  &\ \underbar{$c$}_0 (\underbar{$\alpha$}_{1z}(\| \thickbar{z}\|) + \lambda_{\min} (P) \norm{\thickbar{\eta}}^2  + \norm{e}^2)   \\
     V(\chi) \leq  &\ \thickbarBig{c}_0( \overline{\alpha}_{2z}(\|\thickbar{z}|) + \lambda_{\max} (P) \norm{\thickbar{\eta}}^2 + \norm{e}^2)
\end{aligned}
\end{equation}
where $\underbar{$c$}_0$ and $ \thickbarBig{c}_0$ are positive constants. Consequently, $\alpha_{5}$ and $\alpha_{6}$ can be chosen as
\begin{subequations}
    \begin{align}
    \alpha_{5}(s) & = \underbar{$c$}_0( \underbar{$\alpha$}_{1z}(s) + (\lambda_{\min} (P) + 1) s^2) \\
     \alpha_{6}(s) & = \thickbarBig{c}_0(  \overline{\alpha}_{1z}(s) + (\lambda_{\max} (P) + 1 )s^2)
\end{align}  
\end{subequations}
%
Now consider the flow equations \eqref{eqn: observer flow set1 - transformation} in the set $\mathcal{C}$. 
Recall that, by Assumption \ref{ASS: positive and lower bounded b(w,sigma)}, $b(w, \sigma) $ is lower bounded by the positive constant $b^\star$.
The rate of change of the Lyapunov function $V(\chi)$ evaluated on the flow set is given by 
\begin{align}
    \begin{split}
  \langle \nabla V(\chi), f(\chi) \rangle = & - V(\chi)\exp(T-\tau) \\&
        + \exp(T-\tau) \Big( \langle \nabla  V_1(\thickbarBig{Z}),  f(\thickbarBig{Z}) \rangle  \\& 
        + 2e \big(\thickbar{q}(\thickbar{z}, \thickbar{\eta},e,w, \sigma)  +  b^\star ( -k_p \rho(e) e\\& 
        + \mbox{sat} (\mu(\eta))  - \upvarpi(\bm{\eta^*})\big)  \Big)
    \end{split} \nonumber
\end{align}
Let $ \thickbar{M} = \max(\mu(\eta))$. Then replace the value of $\mu(\eta)$ in the controller by its saturated version,
\begin{align}
    \thickbar{\mu}(\eta) = \thickbarBig{M}\text{sat}\!\left(\frac{\mu(\eta) }{\thickbarBig{M}}\right).
\end{align}
%
%
%
The dynamics of $V(\chi)$ on the flow set can be rewritten as
\begin{align}
    \begin{split}
  \langle \nabla V(\chi), f(\chi) \rangle = & - V(\chi)\exp(T-\tau)\\&
        + \exp(T-\tau) \Big( \langle \nabla  V_1(\thickbarBig{Z}),  f(\thickbarBig{Z}) \rangle  \\& 
        + 2e\thickbar{q}(\thickbar{z}, \thickbar{\eta},e,w, \sigma)  - 2b^\star k_p \rho(e) e^2 \\& 
        + 2b^\star e (\thickbar{\mu}(\eta)  - \upvarpi(\bm{\eta^*})) \Big). 
    \end{split} \nonumber
\end{align}
Applying Young's inequality on the third and last terms yields
\begin{align}
    \begin{split}
  \langle \nabla V(\chi), f(\chi) \rangle & \leq  \big(\!- V(\chi)
       + \langle \nabla  V_1(\thickbarBig{Z}),  f(\thickbarBig{Z}) \rangle + e^2 \\& 
        \quad\  + \norm{\thickbar{q}(\thickbar{z},e,w, \sigma)}^2   - 2b^\star k_p \rho(e) e^2 +e^2 \\& 
      \quad\  + b^\star \norm{\thickbar{\mu}(\eta)  - \upvarpi(\bm{\eta^*})}^2 \big) \exp(T-\tau) 
    \end{split} \nonumber
\end{align}
Given that $\thickbar{q}(0,0,w,\sigma) = 0$, Lemma 11.1 in \cite{Chen_Huang2015} guarantees the existence of smooth positive functions $\pi_2(\cdot) \geq 1$ and $\gamma_z(\cdot) \geq 1$  such that, for all $\thickbar{z} \in \mathds{R}^{n_z}$, $\thickbar{\eta} \in \mathds{R}^{n_\eta}$, and $e \in \mathds{R}$, the inequality
\begin{equation}\label{EQN: qbar}
    \norm{\Bar{q}(\thickbarBig{Z},e, w, \sigma) }^2 \leq \pi_2(\thickbarBig{Z}) \norm{\thickbarBig{Z} }^2 +  \gamma_z(e)e^2
\end{equation}
holds.
%
%
%
%
Applying \eqref{EQN: qbar} and \eqref{EQN: Lemma proof for flow with eta and z} yields
\begin{align}
    \begin{split}
    \langle \nabla V(\chi), f(\chi) \rangle & \leq  
 \big( \!- V(\chi)  - (\Uppsi(\thickbarBig{Z}) - \pi_2(\thickbarBig{Z}))  \norm{\thickbarBig{Z}}^2  \\
 & \quad\  - ( 2 k_p \rho(e) \bar{b}^* - \gamma_r(e) -  \gamma_z(e) - 2 )e^2   \\
 & 
\quad\  + b^* \norm{ \thickbar{\mu}(\eta) - \upvarpi(\bm{\eta^\star})}^2 \big) \exp(T - \tau)
    \end{split} \nonumber
\end{align}
%
Define $\gamma(e) \coloneqq \gamma_r(e) +  \gamma_z(e)$ and assume $\Uppsi(\thickbarBig{Z}) > \pi_2(\thickbarBig{Z})$. We can then design the function $k_p\rho(e)$ to satisfy the inequality $k\rho(e) \geq \max\{\gamma(e)\} + 2 $. Thus:
\begin{align}
    \begin{split}
   \langle \nabla V(\chi), f(\chi) \rangle \leq &  - \big( V(\chi) + \Uppsi(\thickbarBig{Z})\norm{\thickbarBig{Z}}^2 \\&  + ( 2 k_p \rho(e) b^*e^2 \big)\exp(T - \tau)   \\& 
 + \exp(T - \tau) b^* \norm{\thickbar{\mu}(\eta) - \upvarpi(\bm{\eta^\star})}^2. 
    \end{split} \nonumber 
\end{align}
Rewrite $\eta =\bm{\eta^\star} + \zeta$, where
\begin{equation*}
    \zeta = \thickbar{\eta} + b^{-1}(w,\sigma)) Ne.
\end{equation*}
Under Assumption \ref{Ass: Gaussian Process - Lipschitz}, the $\mathcal{GP}$ posterior mean prediction is Lipschitz continuous. Therefore,
%
%
%
\begin{align}
   \begin{split}
         \norm{ \thickbar{\mu}(\eta)  - \upvarpi(\bm{\eta^\star})}^2 \leq &\ \norm{\thickbar{\mu}(\bm{\eta^\star})  - \upvarpi(\bm{\eta^\star})}^2 \\
         &\ 
         + \norm{\thickbar{\mu}(\bm{\eta}^\star + \zeta)  - \thickbar{\mu}(\bm{\eta^\star})}^2\\
       \leq &\ \norm{\thickbar{\mu}(\bm{\eta^\star})  - \upvarpi(\bm{\eta^\star})}^2 
         + L_\mu \norm{\thickbar{\zeta}}^2. 
    \end{split} \label{EQN: GP norm}
\end{align}
%
%
As a result, $\langle \nabla V(\chi), f(\chi)$ satisfies
\begin{align}
    \begin{split}
   \langle \nabla V(\chi), f(\chi) \rangle \leq&  - \Big( V(\chi) + \Uppsi(\thickbarBig{Z})\norm{\thickbarBig{Z}}^2 \\&  + ( 2 k_p \rho(e) \bar{b}^*e^2 \Big)\exp(T - \tau)   \\
   & 
 + \exp(T - \tau) \bar{b}^*(\norm{\thickbar{\mu}(\bm{\eta^\star})  - \upvarpi(\bm{\eta^\star})}^2 
         \\&
         + L_\mu \norm{\thickbar{\zeta}}^2 ). 
    \end{split} \nonumber 
\end{align}
Using Lemma \ref{LEMMA: Upper bounds for GP} and letting $$C_\mu  \coloneqq  L_\mu \norm{\thickbar{\zeta}}^2  + \sqrt{\beta(\rho)} \sigma_{\mathcal{D_S}} (\eta) + \gamma(\rho)$$ yields
\begin{align}
    \begin{split}
  \langle \nabla V(\chi), f(\chi) \rangle & \leq - \big( V(\chi) + \Uppsi(\thickbarBig{Z})\norm{\thickbarBig{Z}}^2 \\&  \quad\, + ( 2 k_p \rho(e) \bar{b}^*e^2 \big)\exp(T - \tau)   \\& \quad\,
 + \exp(T - \tau) \bar{b}^*C_\mu. 
    \end{split} 
\end{align}
%
%
%
%
$C_\mu$  represents the error originating from the $\mathcal{GP}$ model approximation of the ideal internal model continuous function $\upvarpi(\bm{\eta^\star})$.
Consequently, there exists a class $\mathcal{K}^\infty$ function $\bar{\alpha}_1(\|\chi\|)$ such that the inequality 
\begin{align} \label{EQN: Theorem proof - for flow Lyapunov equation}
    \begin{split}
 \langle \nabla V(\chi), f(\chi) \rangle \leq - \bar{\alpha}_1\|\chi\|^2 +  b^{*}C_\mu   \exp(T)
     \end{split} 
\end{align}
holds.
As a result, the states of the system will enter a ball of radius $$\varepsilon_{\mu}=\bar{\alpha}_1^{-1}(b^{*}C_\mu   \exp(T))$$ centered at zero.
%
%
%
During each jump cycle, the $\mathcal{GP}$ model is updated with  $\mathcal{D_S}$ additional training dataset. With a slight abuse of notation, let $\mu_{n+1}(\eta)$ and $\mu_{n}(\eta)$ denote the former and latter predictive mean, respectively.

Then for all $(\tau, \thickbar{z}, \thickbar{\eta}, e, \hat{u}) \in \mathcal{D}$, we have:
\begin{align}
    \begin{split}
         V(\chi)  =  &\ (V_1(\thickbarBig{Z}) + e^2 + (\mu_{n}(\thickbar{\bm{\eta}}) -   \mathbf{u}^\star)^2)
         \exp(T). 
    \end{split}   \label{EQN: previous GP model}   
    \end{align}
and the Lyapunov function evaluated based on updated $\mathcal{GP}$ model using additional training dataset is 
\begin{align}
    \begin{split}
        V(g(\chi))  =  &\ (V_1(\thickbarBig{Z}) + e^2 + (\mu_{n+1}(\thickbar{\bm{\eta}}) -   \mathbf{u}^\star)^2)
         \exp(T) 
    \end{split}    \label{EQN: future GP model}   
    \end{align}
%
Subtracting \eqref{EQN: previous GP model} from \eqref{EQN: future GP model} yields
%

\begin{align}
    \begin{split}
        V(g(\chi))  -   V(\chi)  =  &\ ((\mu_{n+1}(\thickbar{\bm{\eta}}) -   \mathbf{u}^\star)^2 \\
        & 
        - (\mu_{n}(\thickbar{\bm{\eta}}) -   \mathbf{u}^\star)^2
        ) \exp(T). 
    \end{split}\nonumber       
    \end{align}
By substituting the posterior variance $\sigma^2(\thickbar{\eta})$ in place of the generalization error $\epsilon_D$ as defined in \eqref{EQN: Bayesian generalization error}, we have that $\epsilon_D = \sigma^2(\thickbar{\eta})$, which also results in 
%
\begin{align} \label{EQN: Jump Lyanpunov analysis for theorem}
    \begin{split}
        V(g(\chi))  -   V(\chi)  =  &\ (\sigma_{n+1}^2(\thickbar{\eta}) - \sigma_n^2(\thickbar{\eta})
        ) \exp(T). 
    \end{split}    
\end{align}
%
We show that $\sigma_{n+1}^2(\thickbar{\eta}) < \sigma_n^2(\thickbar{\eta})$. We recall the predictive variance given in \eqref{EQN: GP variance} and let the Gram matrix $\thickbarBig{K} = K + \sigma_n^2 I$. The $(n+1) \times 1$ vector $\bm{K}_*$ can be partitioned into a vector $\mathbf{k}_n$ and a scalar $\mathbf{c}$, i.e.,  $$\bm{K}_*(\thickbar{\eta}) = \begin{bmatrix}
    \mathbf{k}_{n}(\thickbar{\eta}) & \mathbf{c}(\thickbar{\eta}) 
\end{bmatrix}^{\top}$$ with $\mathbf{k}_{n}(\thickbar{\eta}) =\col(\bm{K}(\bm{\eta}, \bm{\eta}_1), \cdots{}, \bm{K}(\bm{\eta}, \bm{\eta}_n)) $ and $\mathbf{c}(\thickbar{\eta}) = \bm{K}(\bm{\eta}, \bm{\eta}_{n+1})$.
Let the $(n+1){\text{th}}$ element be denoted by $\thickbar{\eta}'$. The $(n+1) \times (n+1)$ Gram matrix can be partitioned into four sub-matrices:
\begin{align*}
    \thickbarBig{\bm{K}}_{n+1}  & = 
    \begin{bmatrix}
        Q_n &  \mathbf{k}_n(\thickbar{\eta}') \\
       \mathbf{k}_n^{\top}(\thickbar{\eta}') &  \mathbf{c}(\thickbar{\eta}')
    \end{bmatrix},\;\;
    \thickbarBig{\bm{K}}^{-1}_{n+1}  = 
    \begin{bmatrix}
        \tilde{Q}_n &  \tilde{\mathbf{k}}_n(\thickbar{\eta}') \\
       \tilde{\mathbf{k}}_n^{\top}(\thickbar{\eta}') & \tilde{\mathbf{c}}(\thickbar{\eta}').
    \end{bmatrix}
\end{align*}
The matrix $Q_n$ is the $n \times n$ covariance matrix of a $\mathcal{GP}$ trained with $n$ datasets. The inverse of the Gram matrix $\thickbarBig{\bm{K}}^{-1}_{n+1} $ can be obtained by applying the matrix inversion lemma \cite[Section \textbf{A.3}] {Rasmussen_Williams2006}
where
\begin{align*}
    \tilde{Q}_n & =  Q_n^{-1} + Q_n^{-1}\mathbf{k}_n(\thickbar{\eta}')M^{-1} \mathbf{k}^{\top}_n(\thickbar{\eta}') Q_n^{-1}, &\\
    S & =  \tilde{\mathbf{c}}(\thickbar{\eta}') - \mathbf{k}^{\top}_n(\thickbar{\eta}') Q_n^{-1}  \mathbf{k}_n(\thickbar{\eta}'),   &\\
    \tilde{\mathbf{k}}_n(\thickbar{\eta}')  & = - Q_n^{-1}\mathbf{k}_n(\thickbar{\eta}')  M^{-1},  \\
    \tilde{\mathbf{c}}_n(\thickbar{\eta}') & =  S^{-1}.
\end{align*}
We now proceed to compute the posterior variance $\sigma_{n+1}^2(\thickbar{\eta})$ using \eqref{EQN: GP variance}, which signifies the uncertainty of the $\mathcal{GP}$ model's predictions at a given test point $\thickbar{\eta}$ after incorporating $n + 1$ training data points:
\begin{align}
\begin{split}
 \sigma_{n+1}^2(\thickbar{\eta}) =  &\ \bm{K}(\eta, \eta) - 
    \begin{bmatrix}
        \mathbf{k}_n(\thickbar{\eta}) \\
        \mathbf{c}(\thickbar{\eta})
      \end{bmatrix}^{\top}
       \thickbarBig{\bm{K}}_{n+1}^{-1}
       \begin{bmatrix}
         \mathbf{k}_n(\thickbar{\eta}) \\
        \mathbf{c}(\thickbar{\eta})
      \end{bmatrix}\\
    = &\ \bm{K}(\eta, \eta) - \mathbf{k}_{n}^{\top}(\thickbar{\eta}) Q^{-1} \mathbf{k}_{n}(\thickbar{\eta})  \\
    &\    - \frac{(\mathbf{k}_{n}^{\top}(\thickbar{\eta})Q^{-1} \mathbf{k}_{n}(\thickbar{\eta}') - \mathbf{c}(\thickbar{\eta}') )^2 }{\sigma_n^2(\thickbar{\eta}')}.
\end{split} \nonumber
\end{align}
The first two terms represent the posterior variance, denoted as $\sigma_n^2(\thickbar{\eta})$, for the $\mathcal{GP}$ model obtained using $n$ training data points. Thus, we obtain:
\begin{align*}
   \sigma_{n+1}^2(\thickbar{\eta})  = \sigma_{n}^2(\thickbar{\eta}) - \frac{(\mathbf{k}_{n}^{\top}(\thickbar{\eta})Q^{-1} \mathbf{k}_{n}(\thickbar{\eta}') - \mathbf{c}(\thickbar{\eta}') )^2 }{\sigma_n^2(\thickbar{\eta}')}. 
\end{align*}
Define 
 $$\Upgamma(\thickbar{\eta})  \coloneqq \frac{(\mathbf{k}_{n}^{\top}(\thickbar{\eta})Q^{-1} \mathbf{k}_{n}(\eta') - \mathbf{c}(\eta') )^2 }{\sigma_n^2(\eta')},$$ then:
\begin{align}
   \sigma_{n+1}^2(\thickbar{\eta})  = \sigma_{n}^2(\thickbar{\eta}) - \Upgamma(\thickbar{\eta}).\end{align}
Since $\Upgamma(\thickbar{\eta}) \geq 0$, it follows that 
$$ \sigma_{n+1}^2(\thickbar{\eta}) < \sigma_n^2(\thickbar{\eta}).$$
if $\mathbf{k}_{n}^{\top}(\thickbar{\eta})Q^{-1} \mathbf{k}_{n}(\thickbar{\eta}') > \mathbf{c}(\thickbar{\eta}') $, indicating that the uncertainty in the model predictions decreases as new training data are incorporated. However, if $$\mathbf{k}_{n}^{\top}(\thickbar{\eta})Q^{-1} \mathbf{k}_{n}(\thickbar{\eta}') < \mathbf{c}(\thickbar{\eta}'), $$ then 
$$ \sigma_{n+1}^2(\thickbar{\eta}) \approx \sigma_n^2(\thickbar{\eta})$$ which implies that no further improvement in the $\mathcal{GP}$ model is obtainable. Under these conditions, the Lyapunov function during a jump satisfies
\begin{align} \label{EQN: Jump Lyanpunov analysis for theorem - result - condition 1}
    \begin{split}
        V(g(\chi))  -   V(\chi)  \leq   0
    \end{split}    
    \end{align}
Therefore, the solution $\phi$ of the hybrid system $\mathcal{H}_\mu$, is bounded and converges to $\varepsilon_\mu$ as $t\rightarrow \infty$ and $j\rightarrow \infty$. Thus, we prove that $|\phi(t,j)|_{\mathcal{A}} \leq  \varepsilon_\mu $  as (t, j) $\rightarrow \infty \times \infty$, which implies that the error signal enters the set where $|e| \leq \varepsilon_\mu$. This completes the proof. 
\end{Proof}

\begin{remark} \label{Remark: GP prediction inproves with training data}
    The performance of the $\mathcal{GP}$ model progressively improves as more training data are incorporated. As a result, the radius of ball $C_\mu$, centered around zero and representing the uncertainty or error bound, will continue to shrink. This iterative refinement persists until the radius reaches its minimal value.
\end{remark}



\begin{remark}
    Complete knowledge of the system dynamics is not required, unlike the pioneering work in \cite{Bin-Pauline-Marconi-2021}. Here, we require direct measurement of the regulated variable, and the closed-loop system is stabilized using a high-gain stabilization technique. 
    Finally, the algorithm performance depends on the $\mathcal{GP}$ modelling error which is shown to be bounded and reduces as the number of training data increases. 
\end{remark}

\subsection{Training Data Collection and Algorithm Implementation}

In the initial continuous cycle, no historical data are available to train the $\mathcal{GP}$ model. Therefore, the control law \eqref{eqn: control-law regulator equation} is implemented without adaptation i.e., $u = k_p \rho(e) e$. During the jump event, all collected data, comprising the internal model unit trajectory $$\bm{\eta} = \{\eta(t_1), \cdots, \eta(t_n)\}$$ obtained from the ODE solver and the corresponding control inputs $$\bm{u} = \{u(t_1), \cdots{}, u(t_n)\}$$, are utilized to train the $\mathcal{GP}$ model. 
In subsequent continuous cycles, the trained $\mathcal{GP}$ model is employed for prediction, and the control law is implemented as defined in \eqref{eqn: control-law regulator equation}.
At each subsequent jump event, new data points are appended to the existing training dataset, continuously enriching the $\mathcal{GP}$ model with more information.

Gaussian Process models typically involve a computational complexity of $\mathcal{O}(R^3)$ during training due to the matrix inversion step, where $R$ is the size of the training dataset. Although, all previous datasets can be utilized to update the $\mathcal{GP}$ model in theory, this approach can become computationally prohibitive, especially for complex systems, due to the significant increase in time and memory requirements as $R$ grows. To address this challenge, the training process can be limited to only the most recent data points, defined by a sliding window of size $P$. This strategy ensures that the computational cost remains efficient and manageable, even as new data are continuously incorporated.

\begin{algorithm}
\caption{Sliding Window Update for GP Model}
\label{alg:sliding_window}
\begin{algorithmic}[1]
\REQUIRE Sliding window size $\mathcal{W_S}$, internal model unit series $\eta$, control input series $u$, current simulation data $(\bm{\eta}_k, \bm{u}_k)$, $\mathcal{GP}$ model, number of simulation $T_k$
\ENSURE Updated $\mathcal{GP}$ model and predictions
\STATE Initialize window data $ \bm{\mathcal{D_S}} \gets \emptyset$

\FOR{$k = 1$ to $T_k$}
    \STATE Add $(\bm{\eta}_k, \bm{u}_k)$ to $\bm{\mathcal{D_S}}_k$
    \IF{$k < \mathcal{P}$}
        \STATE $\bm{\mathcal{D_S}}_{k+1} = \bm{\mathcal{D_S}}_k \cup \{(\bm{\eta}_k, \bm{u}_k)\}$
        \STATE Continue accumulating data until the window size $W$ is reached
    \ELSIF{$k \geq \mathcal{P}$}
        \STATE $\bm{\mathcal{D_S}}_{k+1} = \{ \bm{\mathcal{D_S}}_k \cup \{(\bm{\eta}_k, \bm{u}_k)\} \backslash \{(\bm{\eta}_{k-\mathcal{W_S}}, \bm{u}_{k-\mathcal{W_S}})\} \}$
    \ENDIF
    \STATE Normalize $\bm{\mathcal{D_S}}_{k+1}$ and train the $\mathcal{GP}$ model
\ENDFOR
\end{algorithmic}
\end{algorithm}

\section{NUMERICAL AND PRACTICAL EXAMPLE} \label{Simulation example}
%
We present some examples to illustrate the effectiveness of the proposed data-driven regulator.
\subsection{Example 1: Controlled Lorenz system}\label{Example 1: Controlled Lorenz systems}

Consider the generalized Lorenz system problem presented in \cite{LiuandJie_Huang2008}. The system dynamics are given by
\begin{equation}\label{Example: numerical example - data-driven control}
\begin{aligned}
    \Dot{z}_1 & = a_{11} z_1 + a_{12} y \\
    \Dot{z}_2 & = a_{3} z_2 + z_1y  \\
    \Dot{y} & = z_1 (a_{21} - z_2) + a_{22}y +  u \\
    e  & = y - w_1
\end{aligned}
\end{equation}
and the exogenous signal is given by
\begin{equation} \label{Example: Exosystem}
    \begin{aligned}
        \dot{w}_1 & = \sigma w_2, \\
        \dot{w}_2 & = - \sigma w_1
    \end{aligned}
\end{equation}
where $(z_1,z_2)$ and $y$ are the state, $e$ is the regulated output, $a = \col(a_{11}, a_{12},a_{21},a_{22}, a_3 )$ are constant parameters satisfying $a_{11} < 0$, $a_3 < 0$. $a$. The parameter $a$ is allowed to undergo some perturbation: 
$$a = (\bar{a}_{11}, \bar{a}_{12},\bar{a}_{21},\bar{a}_{22}, \bar{a}_3 ) + (m_1, m_2, \cdots{}, m_5), $$
where $m = (m_1, m_2, \cdots{}, m_5)$ is an uncertain parameter and $(\bar{a}_{11}, \bar{a}_{12},\bar{a}_{21},\bar{a}_{22}, \bar{a}_3 )$ is the nominal value of $a$. 

It can be verified that the composite system composed of system \eqref{Example: numerical example - data-driven control}  and \eqref{Example: Exosystem} satisfies Assumptions \ref{ASS: positive and lower bounded b(w,sigma)}--\ref{ASS: exosystem input satisfies ISS}. Under Assumption \ref{ASS: exosystem input satisfies ISS}, it is straightforward to show that the steady-state states $\mathbf{y}(w,m, \sigma)$, $\mathbf{z}_1(w, m,\sigma)$, and $\mathbf{z}_2(w, m,\sigma)$ are 
\begin{align*}
 \mathbf{z}_1(w,m, \sigma) & = r_{11} w_1 + r_{12}w_2, \quad \mathbf{y}(w,m, \sigma) = w_1,  \\
  \mathbf{z}_2(w,m, \sigma) &= r_{21}w_1^2 + r_{22}w^2 + r_{23}w_1w_2,
\end{align*}
where
\begin{align*}
    r_{11}(m, \sigma) &= -\frac{a_{11}a_{12}}{\sigma^2 + a_{11}}, \;\;\;\; r_{12}(m, \sigma) = -\frac{a_{12} \sigma }{\sigma^2 + a_{11}^2}, \\
    r_{22}(m, \sigma) & = -\frac{\sigma}{a_{3}} r_{23}, \;\;\;\;\;\;\;\;\;r_{23}(m, \sigma) = -\frac{r_{12}a_3 + 2\sigma r_{11} }{4\sigma^2 + a_{3}^2},\\
     r_{21}(m, \sigma) &= -\frac{a_{3}^2 r_{11} - a_3 \sigma r_{12} + 2 \sigma^2 r_{11} }{a_{3}(a_3^2 + 4 \sigma^2)}. 
\end{align*}
Finally, the ideal feedforward control input $\mathbf{u}(w, \sigma)$ presented in system \eqref{eqn: ideal steady-state control input} is given as
\begin{align}
    \begin{split}
  \mathbf{u}(w,m, \sigma)       = &\ r_{31} w_1 + r_{32}w_2 + r_{33} w_1^3  + r_{34}w_2^3 \\& + r_{35}w_1^2 w_2 
         + r_{36}w_1 w_2^2
    \end{split} \label{Example: ideal control input - data driven control}
\end{align}
where
\begin{align*}
    r_{31}(m,\sigma) & = - b^{-1}(a_{22} + a_{21}r_{11}) \\
    r_{32}(m,\sigma) & =  b^{-1}(\sigma - a_{21}r_{12}) \\
    r_{33}(m,\sigma) & =  b^{-1} r_{11}r_{21}, \quad r_{34}(m,\sigma) = b^{-1} r_{12}r_{22} \\
    r_{35}(m,\sigma) & =  b^{-1}( r_{12}r_{21} + r_{11}r_{23}) \\
    r_{36}(m,\sigma) & =  b^{-1}(r_{11}r_{22} + r_{12}r_{23}).
\end{align*}
There is no analytical approach for obtaining the ideal control input \eqref{Example: ideal control input - data driven control}. We now solve the robust output regulation problem using the proposed regulator. 
%
We perform the simulation with $k_p = 500, 700$, $\rho(e) = e^2 + 1$ and set $\mu(\cdot)$ saturation limit as $\mbox{sat} = 100$.
$n_w = 2$, $n_z = 2$. Thus, $n_\eta = 2(n_w + n_z + 1) = 10$ and 
\begin{align*}
M &= \begin{bmatrix}
    -1 & 1 & 0 & 0 & 0 & 0 & 0 & 0 & 0 & 0\\
    0 & -1 & 1 & 0 & 0 & 0 & 0 & 0 & 0 & 0 \\
    0 & 0 & -1 & 1 & 0 & 0 & 0 & 0 & 0 & 0 \\
    0 & 0 & 0 & -1 & 1 & 0 & 0 & 0 & 0 & 0\\
    0 & 0 & 0 & 0 & -1 & 1 & 0 & 0 & 0 & 0\\
    0 & 0 & 0 & 0 & 0 & -1 & 1 & 0 & 0 & 0 \\
    0 & 0 & 0 & 0 & 0 & 0 & -1 & 1 & 0 & 0 \\
    0 & 0 & 0 & 0 & 0 & 0 & 0 & -1 & 1 & 0\\
    0 & 0 & 0 & 0 & 0 & 0 & 0 & 0 & -1 & 0 \\
     0 & 0 & 0 & 0 & 0 & 0 & 0 & 0 & 0 &- 1
\end{bmatrix}\\ 
  N & = \begin{bmatrix}
   0& 0 & 0 &0 &0 &0 &0 &0 &0 & 1
\end{bmatrix}^{\top}
\end{align*}

Note that \textbf{no adaptation} is performed with the GP model at the first flow event.
The hybrid clock periodic interval is set as $T = 0.1$.
%
We performed the simulation with $\sigma = 0.8$, $a = \col(-10, 10, 28, -1, -2.6667)$. A sliding window size of $\mathcal{P} = 10$. Initial conditions are chosen to be $z(0) = \col(2, -1.8)$, $y(0) = -1.5$, $\eta(0) = 0$, and $w(0) = \col(0,4)$. 
\begin{figure}[ht]
  \centering\setlength{\unitlength}{0.65mm}
  \includegraphics[width=0.5\textwidth]{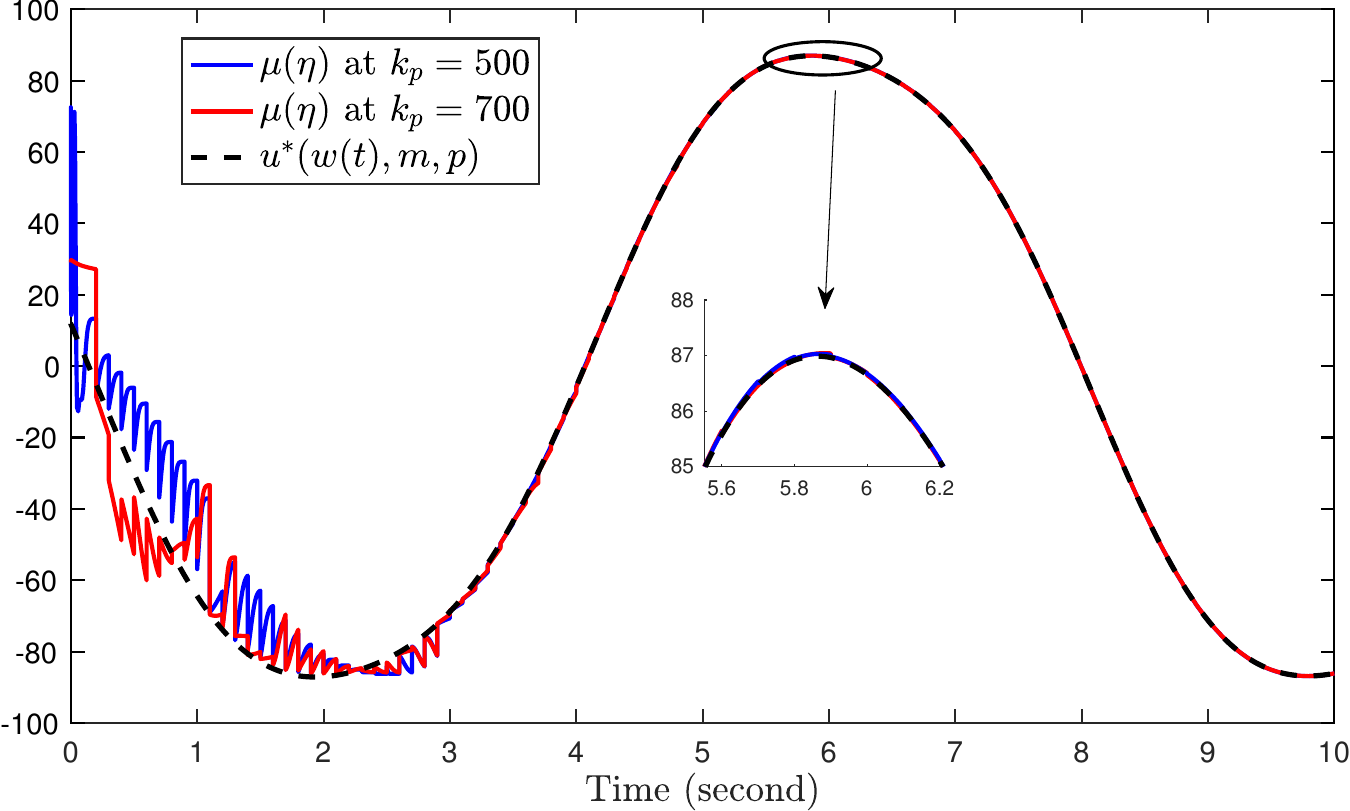}
  \caption{Example \ref{Example 1: Controlled Lorenz systems}: The trajectories of the ideal steady-state control input $\mathbf{u}^\star(w,\sigma, \sigma)$ and its approximation obtained from the  $\mathcal{GP}$ model $\mu(\eta)$ for different values of $k_p$. } \label{Fig: Internal Model trajectory}
\end{figure}
\begin{figure}[ht]
  \centering\setlength{\unitlength}{0.65mm}
  \includegraphics[width=0.5\textwidth]{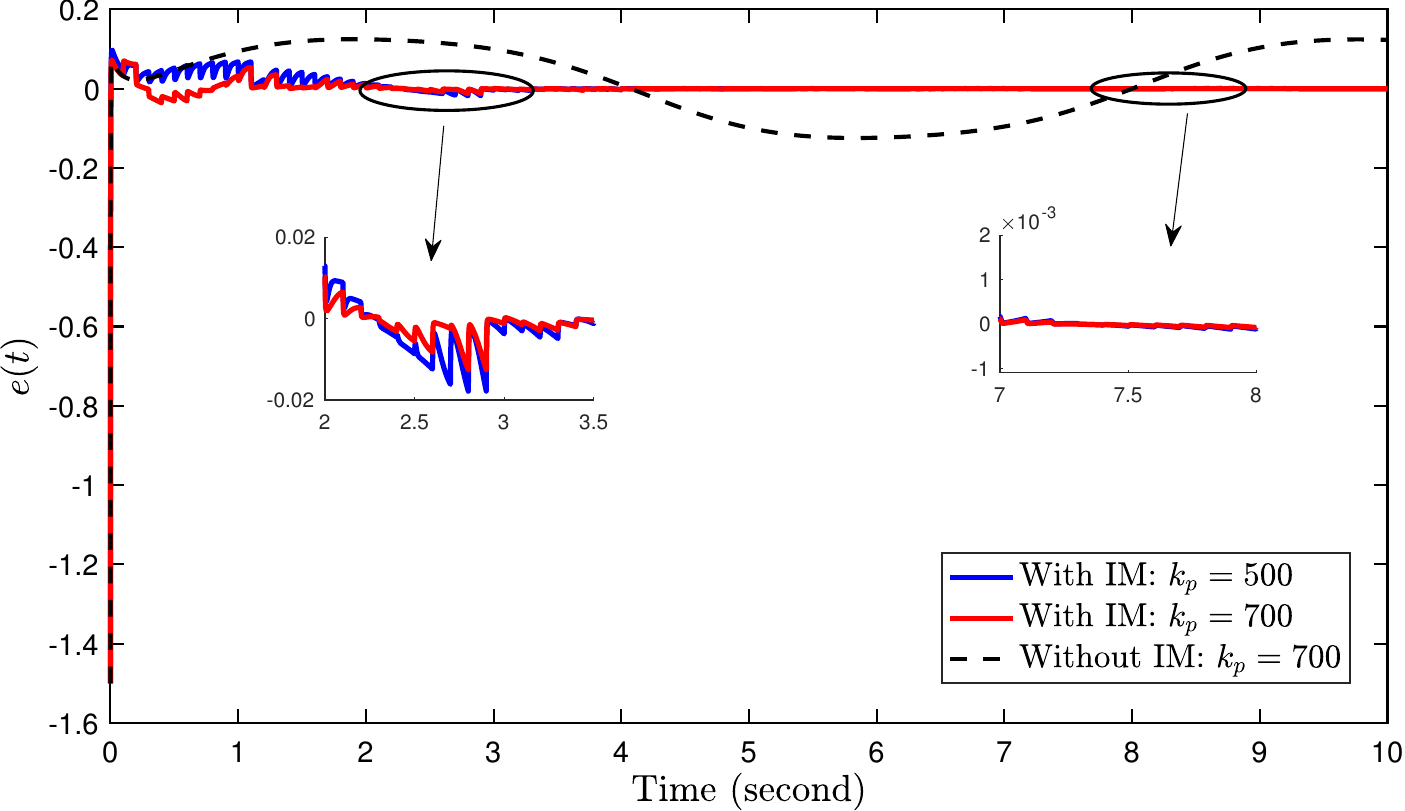}
  \caption{Example \ref{Example 1: Controlled Lorenz systems}: Tracking error trajectories with and without internal model (IM), namely with and without $\mu(\eta)$ in the control law \eqref{eqn: control-law regulator equation} at different $k_p$.} \label{Fig: error_trajectory}
\end{figure}
%
Figure \ref{Fig: Internal Model trajectory} illustrates the trajectories of the ideal feedforward control input, $\mathbf{u}^\star(w,\sigma, \sigma)$, and its estimate, $\mu(\eta)$, generated by the proposed regulator for varying control gains, $k_p$. Figure \ref{Fig: error_trajectory} compares the error trajectory with and without an internal model. The results demonstrate that the approximate internal model, $\mu(\eta)$, exhibits poor performance initially (for the first 4 seconds) but significantly improves with the inclusion of more training data, consistent with Remark \ref{Remark: GP prediction inproves with training data}. Furthermore, the regulator's performance increases with higher control gains, $k_p$. The saturation function effectively mitigates the peaking associated with the high gains. 
\subsection{Example 2: System with a nonlinear exosystem}\label{Example: Typical nonlinear system }
\begin{figure}[ht]
  \centering\setlength{\unitlength}{0.65mm}
    \includegraphics[width=0.5\textwidth]{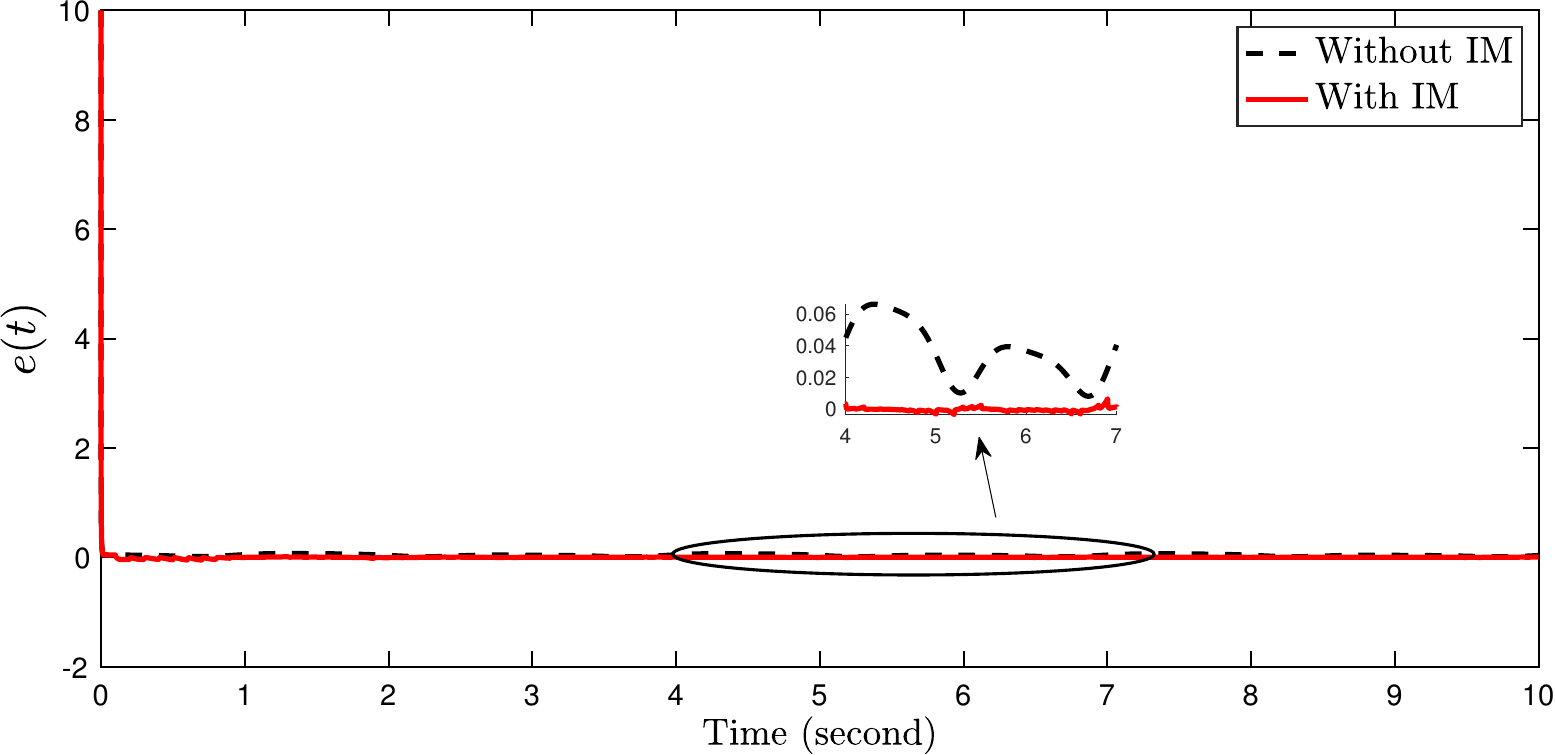}
  \caption{Example \ref{Example: Typical nonlinear system }: Tracking error trajectories with and without internal model (IM), namely with and without $\mu(\eta)$ in the control law \eqref{eqn: control-law regulator equation}.} \label{Fig: Example 2 error_trajectory}
\end{figure}

We consider the nonlinear system taken from \cite{Pauline_Bin_Marconi2020}
\begin{equation} \label{EQN: Example2: nonlinear system}
    \begin{aligned}
        \Dot{z}  = & -2z + y + 2w_1 \\
        \dot{y}  = &\  w_2^2 + z y + u \\
        e =  &\ y - w_1
    \end{aligned}
\end{equation}
The dynamics of the exosystem are described by the nonlinear differential equation:
\begin{equation}
    \begin{aligned}
        \dot{w}_1 & = w_2 \\
        \dot{w}_2 & = - w_1 - w_1^3
    \end{aligned}
\end{equation}
System \eqref{EQN: Example2: nonlinear system} conforms to the structure of \eqref{eqn: general nonlinear system dynamics}, and, as such, falls within the scope of the analysis presented in this article. The simulation is performed with $k_p = 300$, the $\mu(\cdot)$ saturation limit is set to $\mbox{sat} = 25$ and $\rho(e) = e^2 + 1$. We choose $n_\eta = 2(n_w+1) = 6$ and
\begin{align*}
M &= \begin{bmatrix}
    -1 & 1 & 0 & 0 & 0 & 0  \\
    0 & -1 & 1 & 0 & 0 & 0 \\
    0 & 0 & -1 & 1 & 0 & 0  \\
    0 & 0 & 0 & -1 & 1 & 0\\
    0 & 0 & 0 & 0 & -1 & 1 \\
    0 & 0 & 0 & 0 & 0 & -1  
\end{bmatrix}\\ 
  N & = \begin{bmatrix}
    0 &0 &0 &0 &0 & 1
\end{bmatrix}^{\top}
\end{align*}
The simulation is performed with initial conditions $(z(0),y(0)) =\col(1,10)$ and $w(0) =\col(0,4)$. The error trajectories with and without an internal model are illustrated in Figure \ref{Fig: Example 2 error_trajectory}.
\subsection{Example 3: Control of a continuous fermenter }\label{Subsection: continuous bioreactor example}
\begin{figure}\centering
\begin{tikzpicture}
\node[inner sep=0pt] (bioreactor) at (1,0)
    {\includegraphics[width=.15\textwidth]{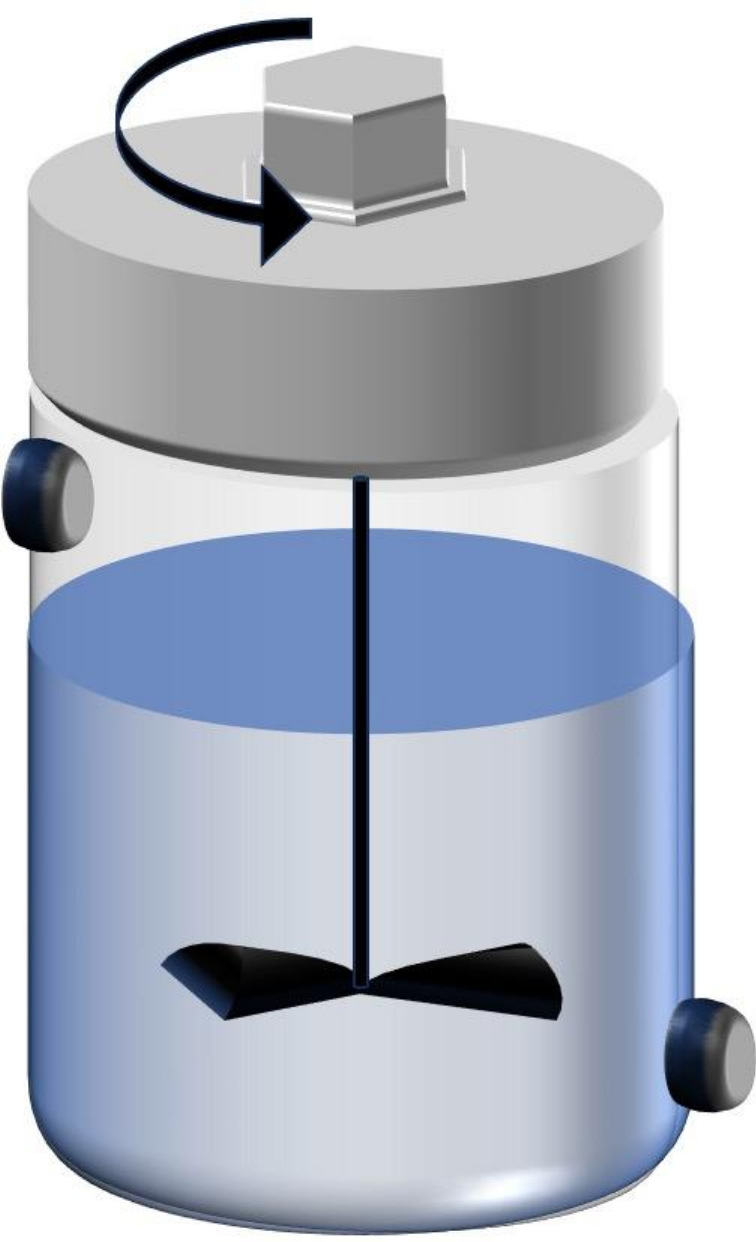}};
      \draw[{Stealth[length=3mm]}-] (bioreactor.160) -- ++ (-1.5,0) 
    node[near end](Inlet){} node[above]{$D, S_f$};
    \draw[-{Stealth[length=3mm]}] (bioreactor.311) -- ++ (1.5,0) 
    node[near end](outlet){} node[midway,above]{$D$}
    node[near end](outlet2){} node[near end, below]{$X, S, P$};
\end{tikzpicture}
 \caption{Continuous bioreactor.}
    \label{Fig: bioreactor schematics}
\end{figure}

Consider the anaerobic continuous bioreactor in Figure \ref{Fig: bioreactor schematics}. Assume that the bioreactor is perfectly mixed and the volume $V$ due to the influent and effluent flow rates $F$ being equal. $S_f$ is the influent substrate concentration, and $X$, $S$, and $P$ are the cellular biomass concentration, the growth limiting substrate concentration, and the product concentration, respectively.  
For ethanol production,  $X$, $S$, and $P$ are the yeast, glucose, and ethanol concentrations, respectively, and $D$ is the dilution rate.
%
The bioreactor in Figure \ref{Fig: bioreactor schematics} can be described by the nonlinear equations \cite{HENSON_Seborg_1992}.
%
%

\begin{equation}\label{EQN: Continuous bioreactor model with three state}
    \begin{aligned}
      \dfrac{\mathrm{d} X}{\mathrm{d} t} &= -D X +\mu(S,P) X,\\
      \dfrac{\mathrm{d} S}{\mathrm{d} t} &= D( S_f - S)- \frac{ \mu(S,P) X}{\gamma_{X/S} } \\
    \dfrac{\mathrm{d} P}{\mathrm{d} t} &= -D P + \big[\alpha\mu(S,P)+\beta\big] X  
    \end{aligned}
\end{equation}
where $\mu(S,P)$ is the specific growth rate, $\gamma_{X/S} $ is the cell mass yield, and $\alpha$ and $\beta$ are product yield parameters. 
The signal $e$ is the regulated output.
The growth rate function,
\begin{align}
    \mu(S,P) = \frac{\mu_m \Big(1 - \frac{P}{P_m} \Big)S }{K_m + S + \frac{S^2}{K_i}},
\end{align}
accounts for both production and substrate inhibition, 
where $\mu_m$, $P_m$, $K_m$, and $K_i$ are the maximum specific growth rate, production saturation constant, substrate saturation constant, and substrate inhibition constant, respectively. The cell-mass yield $\gamma_{X/S}$ and maximum specific growth rate $\mu_m$ are sensitive to variations in operating conditions and are considered \emph{measurable disturbances} due to their potential for significant time-varying behaviour \cite{wang2024nonparametric,HENSON_Seborg_1992}. 
Assume that the disturbance is generated by the signal
\begin{equation}\label{EQN: cell-mass yield disturbance}
    \gamma_{X/S} = \thickbar{\gamma}_{X/S} + a_0 \sin(\omega t).
\end{equation}
Nearly optimum operating conditions and parameters used for model \eqref{EQN: Continuous bioreactor model with three state} are in Table \ref{Table: continuous bioreactor process parameters and initial values}.

\begin{figure}[ht]
  \centering\setlength{\unitlength}{0.65mm}
   \includegraphics[width=0.5\textwidth]{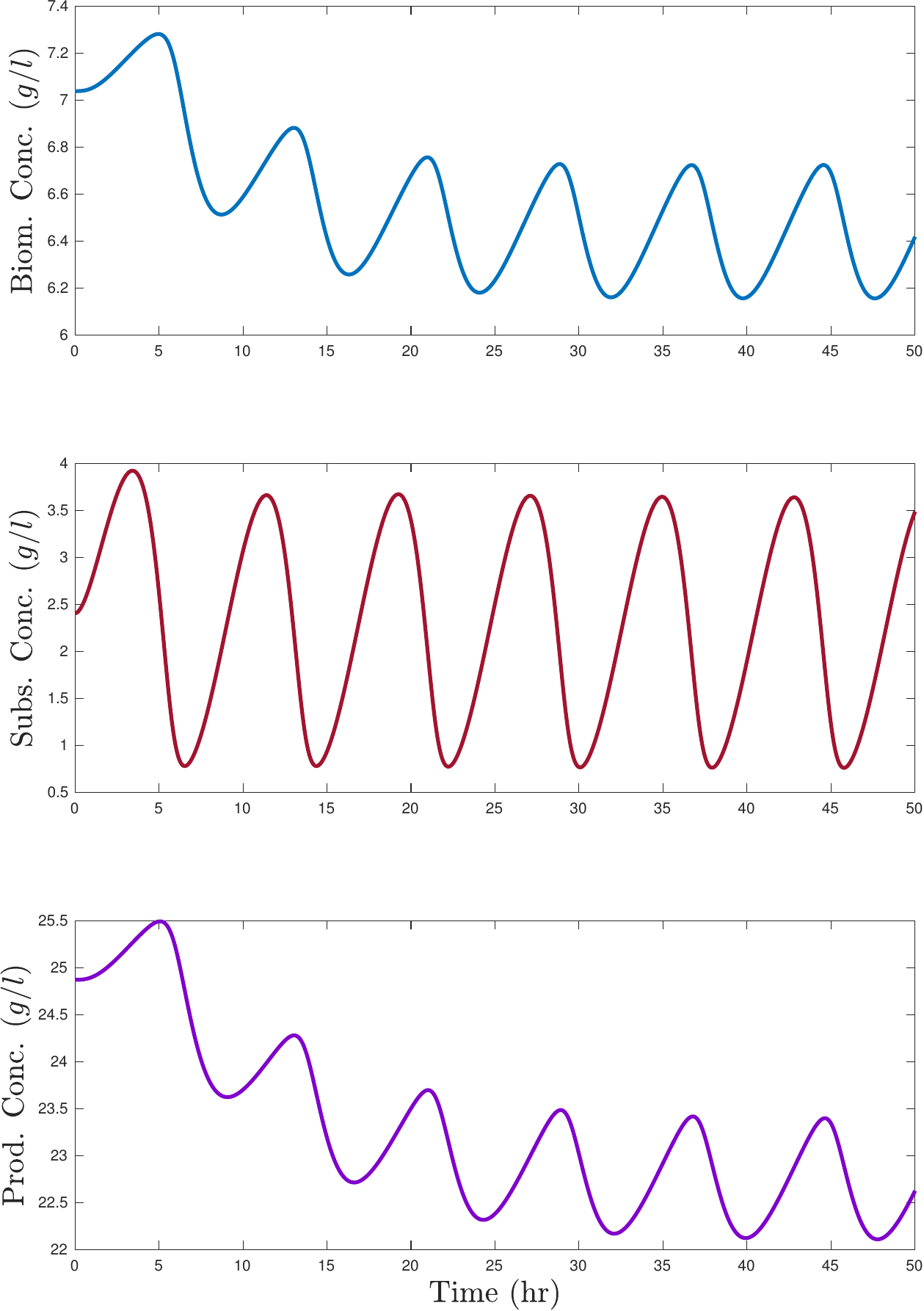}
  \caption{Example \ref{Subsection: continuous bioreactor example}: Dynamic responses of the continuous bioreactor model for a sinusoidal disturbance in the cell-mass yield $\gamma_{X/S} $ represented by equation \eqref{EQN: cell-mass yield disturbance}. The three subplots show, from top to bottom, the trajectories of biomass concentration $X$, substrate concentration $S$ and product concentration $P$.  } \label{Fig: Bioreactor open loop graph}
\end{figure}

\begin{table}[ht]
  \caption{Process Parameters and Initial values}
  \label{Table: continuous bioreactor process parameters and initial values}
  \centering
\begin{tabular}{l c c} 
    \toprule
\thead{Process parameters\\ variables}
    &   \thead{Nominal Value} & Initial Condition \\ 
    \midrule
$\thickbar{\gamma}_{X/S}$  (\unit{g \:g^{-1} }) &   0.4 & \textendash \\
$\alpha $ (\unit{g \: g^{-1}}) &   2.2 & \textendash\\
$\beta $ (\unit{h^{-1}}) &   0.2 & \textendash \\
$\mu_m $ (\unit{h^{-1}}) &   0.48 &\textendash \\
$P_m$ (\unit{g \: l^{-1}})  &   50 & \textendash \\ 
$K_m$ (\unit{g \: l^{-1}})  &   1.2 & \textendash \\ 
$K_i$ (\unit{g \: l^{-1}})  &   22 & \textendash \\ 
$D$ (\unit{h^{-1}})  &  0.15 & \textendash \\ 
$S_f$ (\unit{g \: l^{-1}})  &   20 & \textendash \\ 
$X(0)$ (\unit{g \: l^{-1}}) &   \textendash & 7.038 \\ 
$S(0)$ (\unit{g \: l^{-1}}) &   \textendash & 2.404 \\ 
$P(0)$ (\unit{g \: l^{-1}}) &   \textendash & 24.87 \\ 
    \bottomrule
\end{tabular}
\end{table}
%
The product $P$ is affected by the dilution rate $D$ and some external disturbances. 
The substrate feed concentration $S_f$ is the manipulated input $u$. The term $\mu_m$ is treated as an uncertain parameter, the cell mass yield $\gamma_{X/S}$ is the disturbance, and $X$, $S$, and $P$ are the process state variables.
The open-loop simulation for parameters $a_0 = 0.2$ and $\omega = 0.8$ is shown in Figure \ref{Fig: Bioreactor open loop graph}. Understanding both the system dynamics and the exosystem can significantly aid in designing the model set, particularly by incorporating constraints such as saturation limits. This is especially important as the control input is defined as the substrate input concentration, $S_f$. Therefore, the control input is bounded within the range $u = S_f \in [0,45]$.  The control law is given by $$u = - k_p(S - S_p) + \mbox{sat}(\mu(\eta)),$$ where $S_p$ is the set point for the substrate concentration. The closed-loop simulation is performed with $S_p = 23.4$, $k_p = 30$ for $n_\eta = 6$ and a sliding window $\mathcal{P} = 10$. 

\begin{figure}[ht]
  \centering\setlength{\unitlength}{0.65mm}
  \includegraphics[width=0.5\textwidth]{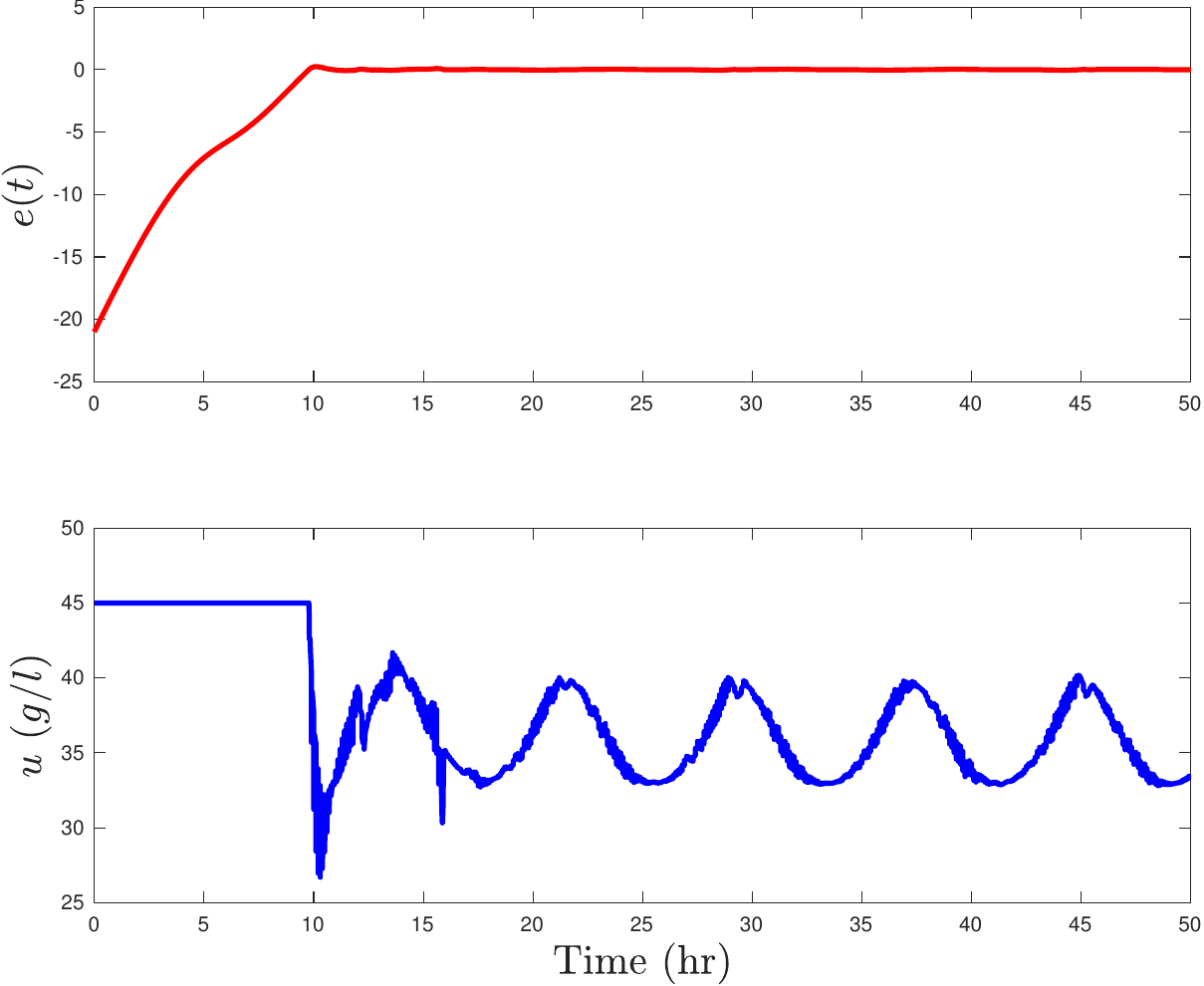}
  \caption{Example \ref{Subsection: continuous bioreactor example}: The upper plot illustrates the trajectory of the tracking error, $e$, while the lower plot depicts the evolution of the control input, $u$, over time.} \label{Fig: Bioreactor closed-loop error trajectory}
\end{figure}

\begin{figure}[ht]
  \centering\setlength{\unitlength}{0.65mm}
   \includegraphics[width=0.5\textwidth]{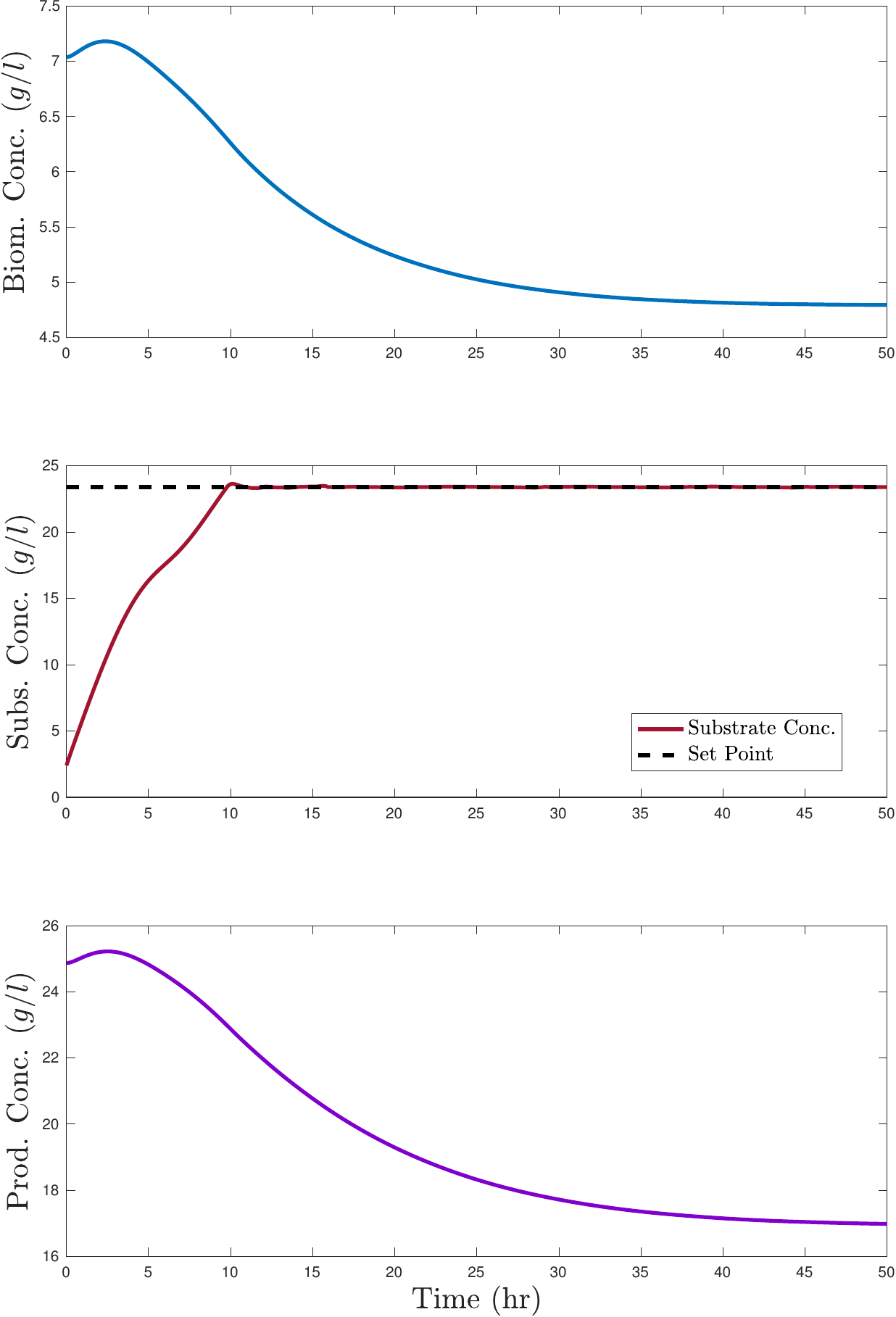}
  \caption{Example \ref{Subsection: continuous bioreactor example}: Closed-loop dynamic responses of the continuous bioreactor model for a sinusoidal disturbance in the cell-mass yield $\gamma_{X/S} $. The three subplots show, from top to bottom, the trajectories of biomass concentration $X$, substrate concentration $S$ and product concentration $P$.} \label{Fig: Bioreactor closed loop graph of all states}
\end{figure}

Figure \ref{Fig: Bioreactor closed-loop error trajectory} illustrates the error trajectory alongside the evolution of the control input over time. The stabilization and saturation functions effectively stabilize the system within the neighbourhood of the origin, while the approximate internal model, $\mu(\eta)$, ensures that the system is regulated to achieve the desired objective. Figure \ref{Fig: Bioreactor closed loop graph of all states} shows the closed-loop dynamic responses of the continuous bioreactor model for a sinusoidal disturbance in the cell-mass yield $\gamma_{X/S} $.


\section{Conclusion}
\label{Conclusion}
This article presents a hybrid adaptive data-driven regulator for a class of nonlinear systems. The proposed control algorithm utilizes datasets generated during continuous-time events to construct a data-driven model that approximates the unknown steady-state continuous nonlinear map online using supervised learning techniques. Numerical and theoretical analyses demonstrated a decreasing prediction error of the regression model as the size of the training dataset increases.
Additionally, through rigorous analytical analysis, we showed that the closed-loop system is stable and converges to a compact set, with the size of this set diminishing as the training dataset grows. The proposed method does not rely on complete system knowledge, making it robust and well-suited for systems with modelling errors or uncertainties. Future work will focus on extending the framework to systems with unknown optimal operating points, further enhancing its applicability.

\section*{Acknowledgments}
The authors would like to express their sincere gratitude to Michelangelo Bin for his help while writing this paper.

\section{Appendix} \label{Appendix}

\subsection{The Proof of Lemma \ref{COR: ISS results for z dynamics system}} \label{Appendix: Proof of Corollary - ISS of Z subsystem}

We first show that condition \eqref{EQN: Lyapunov condition for Lemma} is fulfilled with
\begin{equation}
\begin{aligned}
    V_1(\thickbarBig{Z}) \geq  &\ \underbar{$\alpha$}_{1z}(\| \thickbar{z}\|) + \lambda_{\min} (P) \norm{\thickbar{\eta}}^2 \nonumber \\
     V_1(\thickbarBig{Z}) \leq  &\ \overline{\alpha}_{2z}(\|\thickbar{z}|) + \lambda_{\max} (P) \norm{\thickbar{\eta}}^2 
\end{aligned}
\end{equation}
We see that $\underbar{$\alpha$}_1 $ and $\thickbar{\alpha}_1$ can be taken as
\begin{subequations}
    \begin{align}
    \underbar{$\alpha$}_1(s) & =  \underbar{$\alpha$}_{1z}(s) + \lambda_{\min} (P) s^2  \\
     \thickbar{\alpha}_1(s) & =  \overline{\alpha}_{1z}(s) + \lambda_{\max} (P) s^2 
\end{align}  
\end{subequations}
We now consider the flow system: 

Under Assumption \ref{ASS: exosystem input satisfies ISS}, given a compact set $\Upupsilon \subset \mathds{R}^{n_w} \times \mathds{M} \times \Sigma $, there exists a smooth function $V_{\thickbar{z}}: \Upupsilon \rightarrow \mathds{R}_{\geq 0} $, that satisfies $\underbar{$\alpha$}_{\bar{z}}(\norm{\thickbar{z}}) \leq V_{\thickbar{z}}(\thickbar{z}) \leq \overline{\alpha}_{\bar{z}}(\norm{\thickbar{z}})$ for some class $\mathcal{K}_\infty$ functions $\underbar{$\alpha$}_{\bar{z}}$, $\overline{\alpha}_{\bar{z}}$. Furthermore, there exists a known positive definite function $\rho(e)$ such that, for any $\varphi \in \Upupsilon$,
   \begin{align} \label{EQN: z-subsystem lyapunov function during flow}
        \langle \nabla V_{\thickbar{z}}(\thickbar{z}), f_z (\thickbar{z}, e)\rangle &\leq - \alpha_{\bar{z}} (\norm{\thickbar{z}}) + \delta \rho(e)
    \end{align}
along the trajectories of the $\thickbar{z}$ subsystem in system \eqref{EQN: augmented system}, 
    where $\alpha_{\bar{z}}(\cdot)$ satisfies 
    $\lim_{s \rightarrow 0^+} \sup(\alpha_{\bar{z}}^{-1}(s^2)/s ) < +\infty$.
%
    Then, by the changing of supply rate technique \cite{Sontag-Teel-1995}, given any smooth function $\Omega_1(\thickbar{z}) > 0 $, there exist a $C^1$ function $V_{1z}(\thickbar{z}) $ and some class $\mathcal{K}_\infty$ functions $\underbar{$\alpha$}_{1z}(\cdot)$ and $\overline{\alpha}_{1z}(\cdot)$ satisfying 
$$\underbar{$\alpha$}_{1z}(\| \thickbar{z}\|) \leq V_{1z}(\thickbar{z}) \leq \overline{\alpha}_{1z}(\|\thickbar{z}|) $$ such that system \eqref{EQN: z-subsystem lyapunov function during flow} can be rewritten as
\begin{align}\label{EQN: V_1z}
     \langle \nabla V_{1\thickbar{z}}(\thickbar{z}), f (\thickbar{z}, e)\rangle\leq - \Omega_1(\thickbar{z}) \norm{\thickbar{z}}^2 +   \gamma_z(e) e^2 
\end{align}
where $\gamma_z(e)$ is some known positive function.  
Since $M$ is Hurwitz, there exists a positive definite matrix  $P \in \mathds{R}^{n_\eta \times n_\eta} $ satisfying $P M + M^{\top} P = - 2I$. Next, we  pose a Lyapunov function 
\begin{equation}\label{EQN: Lyapunov Function for Z and eta}
    V_1(\thickbarBig{Z}) = V_{1z} +  \thickbar{\eta}^{\top} P \thickbar{\eta}.
\end{equation}

The Lyapunov function $V_1(\thickbarBig{Z})$ evaluated on the flow set subsystem $(\thickbar{z}, \thickbar{\eta}) \in \mathcal{C}$, and substituting for \eqref{EQN: V_1z}, gives:
\begin{align}
    \begin{split}
\langle \nabla V_1(\thickbarBig{Z}), f (\thickbarBig{Z})\rangle    \leq&\ \thickbar{\eta}^{\top} (P M + M^{\top} P ) \thickbar{\eta} +  2\norm{\eta}  \\
&\ \norm{P\thickbar{r}(\thickbar{z},e,w,\sigma)} 
    +  \langle \nabla V_{\thickbar{z}}(\thickbar{z}), f (\thickbar{z}, e)\rangle.
     \end{split} \nonumber 
\end{align}
Substituting \eqref{EQN: V_1z} yields
\begin{align}
    \begin{split}
     \langle \nabla V_1(\thickbarBig{Z}), f (\thickbarBig{Z})\rangle \rangle    \leq& - 2 \|\thickbar{\eta} \|^2 + \norm{\thickbar{\eta}}^2 + \norm{P\thickbar{r}(\thickbar{z},e,w,\sigma) }^2 \\ 
     & -  \Omega_1(\thickbar{z}) \norm{\thickbar{z}}^2 +    \gamma_z(e) e^2 
    \end{split} \nonumber 
\end{align}
Since $\thickbar{r}(0,0,0,\sigma) = 0$, by \cite[Lemma 7.8]{HuangJ-2004},  there exist some smooth positive functions $\pi_1(\cdot) \geq 1$ and $\nu(\cdot) \geq 1$ and a positive constant $c_0 \in \mathds{R}_{\geq 0}$ such that, for all $\thickbar{z} \in \mathds{R}^{n_z}$, $e \in \mathds{R}$, 
\begin{equation}\label{EQN: PR norm}
   \norm{P\thickbar{r}(\thickbar{z},e,w,\sigma)}^2 \leq \pi_1(\thickbar{z} ) \norm{\thickbar{z}}^2 + \nu(e)e^2.
\end{equation}
Substituting for $\norm{P\thickbar{r}(\thickbar{z},e,w,\sigma)}^2$ yields
\begin{align}
    \begin{split}
    \langle \nabla V_1(\thickbarBig{Z}), f (\thickbarBig{Z})\rangle    \leq& - \norm{\thickbar{\eta}}^2  - (\Omega_1(\thickbar{z}) - \pi_1(\thickbar{z}) ) \norm{\thickbar{z}}^2 \\
    &  + (\nu(e) +   \gamma_z(e) ) e^2.
    \end{split} \nonumber
\end{align}
Let $\Omega_1(\thickbar{z}) \geq \pi_1(\thickbar{z})$ and $\gamma_r(e) = \nu(e) +   \gamma_z(e) $. Therefore,
\begin{align}
\begin{split}
     \langle \nabla V_1(\thickbarBig{Z}), f (\thickbarBig{Z})\rangle \leq& - \Omega_1(\thickbar{z}) \norm{\thickbar{z}}^2 - \norm{\thickbar{\eta}}^2 \\
     & 
     + \gamma_r(e)e^2
\end{split} \nonumber \\
    \leq& -\Uppsi(\thickbarBig{Z})\thickbarBig{Z}^2  + \gamma_r(e)e^2,     \label{EQN: Lemma proof for flow with eta and z} 
\end{align}
which is similar to \eqref{EQN: zn subsystem flow equation}. This completes the proof.
%
%




\hfill $\Box$\par

\bibliographystyle{ieeetr}
\bibliography{my_bibliography}
 \begin{IEEEbiography}
 [{\includegraphics[width=1in,height=1.25in, clip,keepaspectratio]{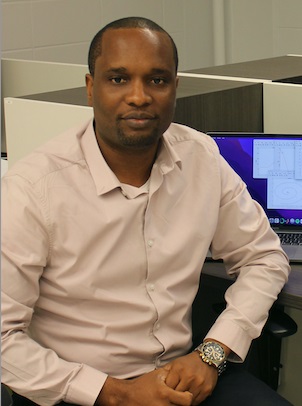}}]{Telema Harry}
 received his B.Tech.\ and M.Sc.\ degrees in Chemical Engineering and Chemical Process Engineering from Rivers State University, Port Harcourt, Nigeria, and University College London, England, respectively. He is currently pursuing a Ph.D. in Chemical Engineering with specialization in Control System Engineering from Queen's University, Kingston, Canada.
 \end{IEEEbiography}

\begin{IEEEbiography}[{\includegraphics[height=1.25in, clip,keepaspectratio]{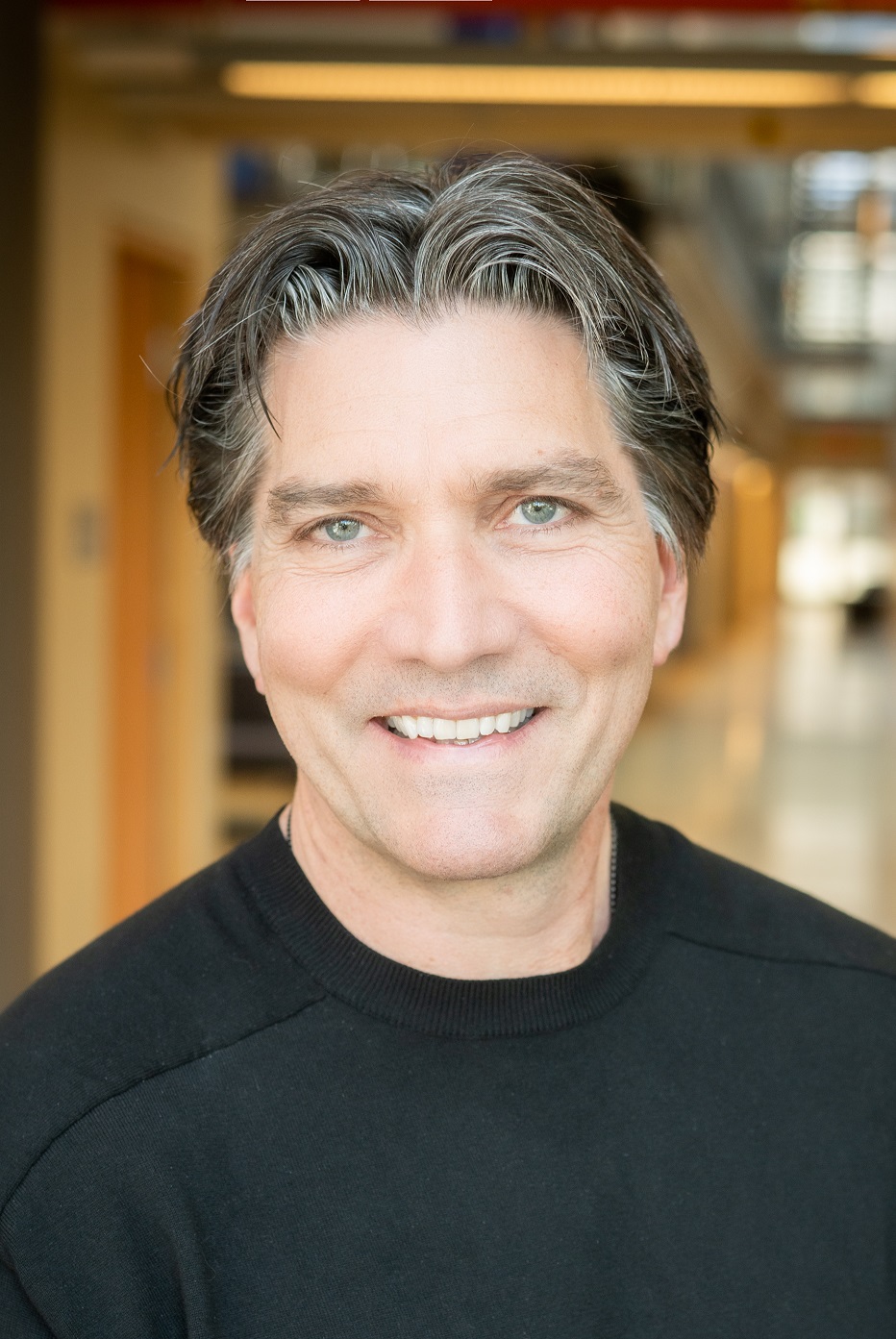}}]{Martin Guay} received a Ph.D. from Queen’s University, Kingston, ON, Canada in 1996. He is a Professor in the Department of Chemical Engineering at Queen’s University. His research interests include nonlinear control systems, especially extremum-seeking control, nonlinear model predictive control, adaptive estimation and control, and geometric control. 

He received the Syncrude Innovation Award, the D.G. Fisher from the Canadian Society of Chemical Engineers, and the Premier Research Excellence Award. He is a Senior Editor of the IEEE Transactions on Automatic Control and the Editor-in-Chief of the Journal of Process Control. He is also an Associate Editor for Automatica and the Canadian Journal of Chemical Engineering.
  \end{IEEEbiography}

\begin{IEEEbiography}[{\includegraphics[width=1in,height=1.25in, clip,keepaspectratio]{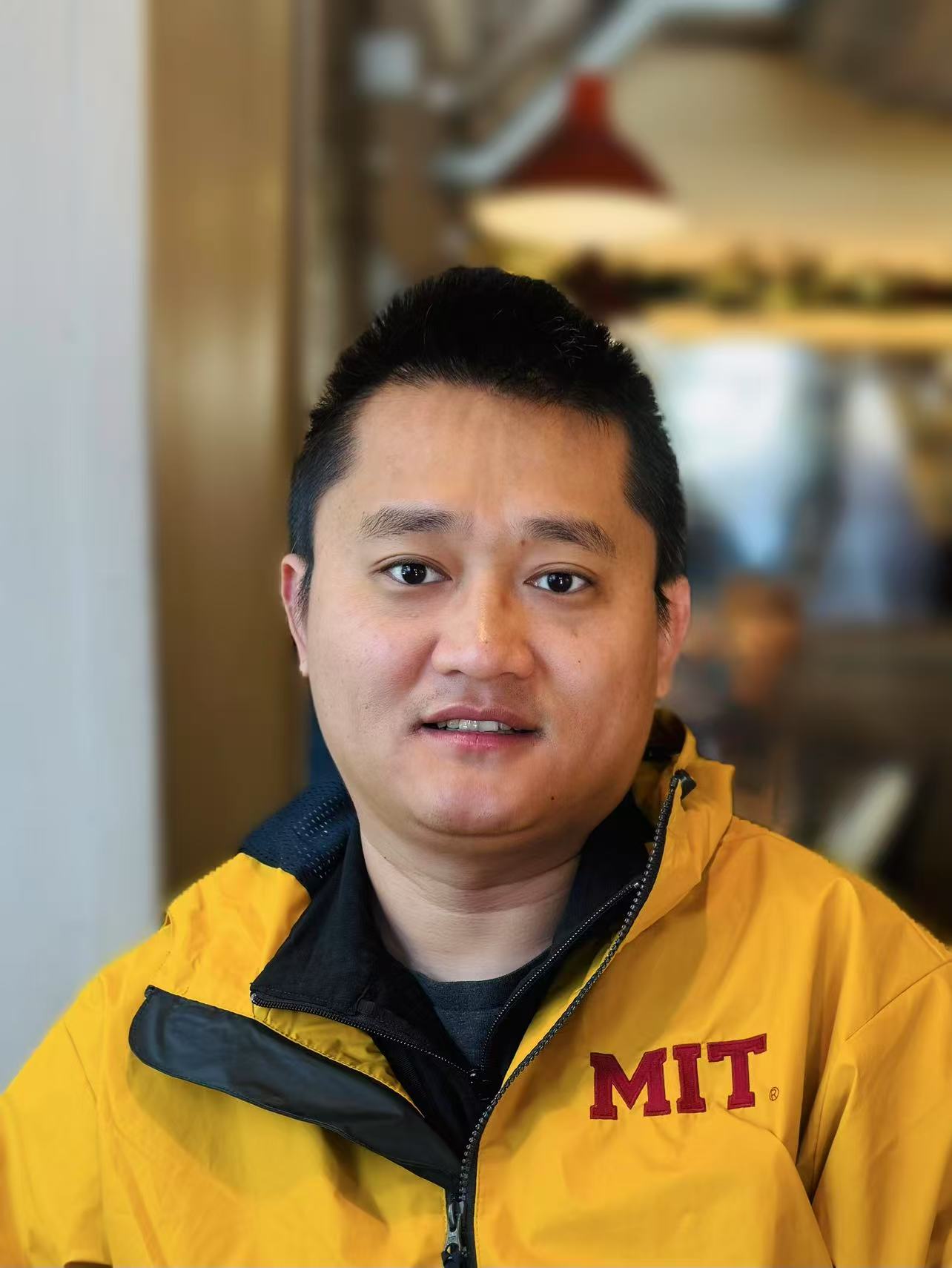}}]{Shimin Wang} 
received a B.Sci.\ in Mathematics and Applied Mathematics and an M.Eng.\ in Control Science and Control Engineering from Harbin Engineering University in 2011 and 2014, respectively. He then received a Ph.D. in Mechanical and Automation Engineering from The Chinese University of Hong Kong in 2019. 
He received the NSERC Postdoctoral Fellowship award in 2022. From 2014 to 2015, he
was an assistant engineer at the Jiangsu Automation
Research Institute, China State Shipbuilding Corporation Limited. From 2019
to 2023, he held postdoctoral positions at the University of Alberta and Queens University. He is currently a postdoctoral associate
at the Massachusetts Institute of Technology.
\end{IEEEbiography}

\begin{IEEEbiography}[{\includegraphics[width=1in,height=1.25in, clip,keepaspectratio]{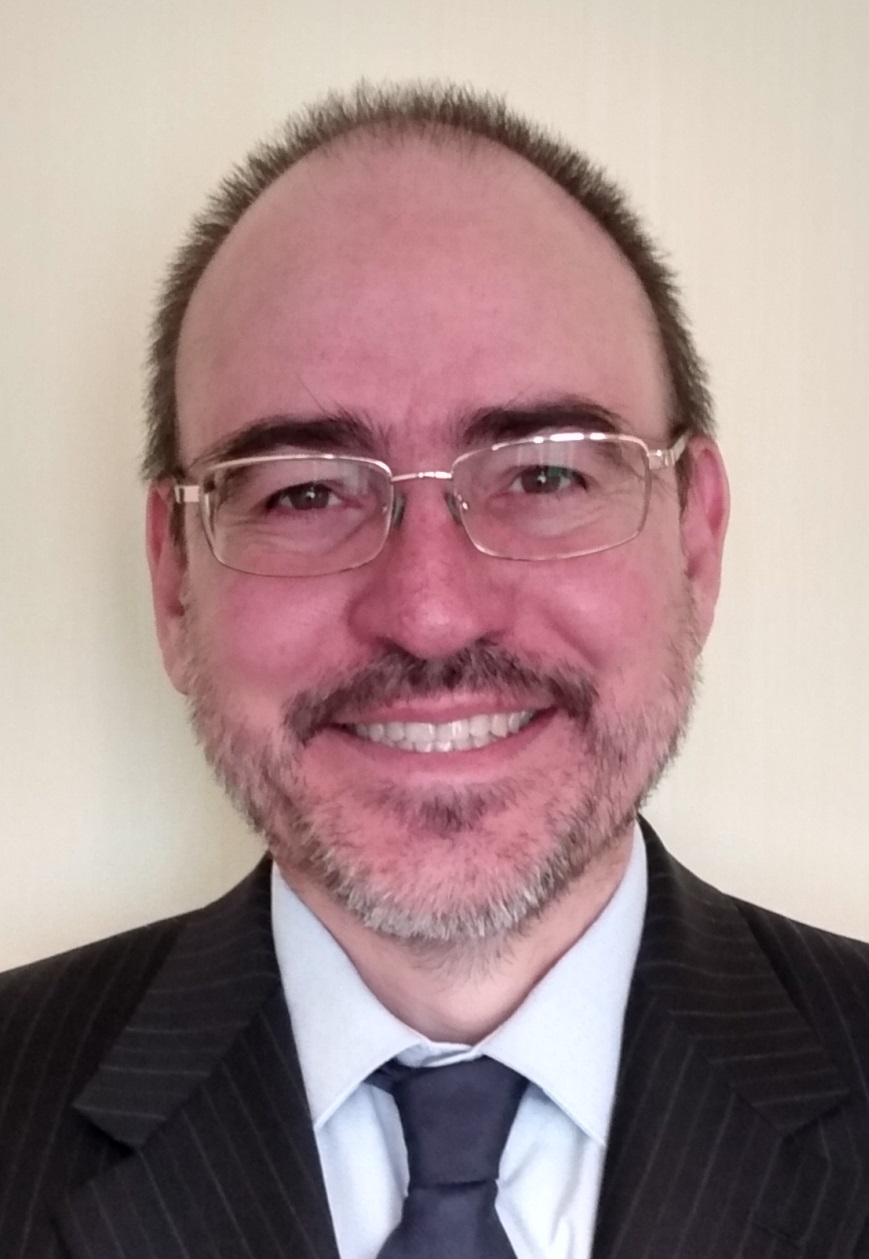}}]{Richard D. Braatz} is the Edwin R. Gilliland Professor at the Massachusetts Institute of Technology (MIT) where he does research in applied mathematics and robust control theory with applications to advanced manufacturing systems. He received an M.S. and Ph.D. from the California Institute of Technology and was on the faculty at the University of Illinois at Urbana–Champaign and a Visiting Scholar at Harvard University before moving to MIT. He is a past Editor-in-Chief of IEEE Control Systems and a past President of the American Automatic Control Council. Honors include the AACC Donald P. Eckman Award, the Curtis W. McGraw Research Award from the Engineering Research Council, the Antonio Ruberti Young Researcher Prize, and best paper awards from IEEE- and IFAC-sponsored control journals. He is a member of the U.S. National Academy of Engineering and a Fellow of IEEE and IFAC.
\end{IEEEbiography}
\end{document}